\documentclass[amsmath,amssymb,12pt,aps]{revtex4-2}
\usepackage[utf8]{inputenc} 
\usepackage{graphicx}
\usepackage{float}
\usepackage{natbib}
\usepackage{wrapfig}
\usepackage{xcolor}
\usepackage{dcolumn}
\usepackage[english]{babel}
\begin{document}
	
\title{Generalized frustration in the multidimensional Kuramoto model}

\author{Marcus A. M. de Aguiar }

\affiliation{Instituto de F\'isica `Gleb Wataghin', Universidade Estadual de Campinas, Unicamp 13083-970, Campinas, SP, Brazil}

\begin{abstract}
	
The Kuramoto model was recently extended to arbitrary dimensions by reinterpreting the oscillators as particles moving on the surface of unit spheres in a D-dimensional space. Each particle is then represented by a D-dimensional unit vector. For $D=2$ the particles move on the unit circle and the vectors can be described by a single phase, recovering the original Kuramoto model. This multidimensional description can be further extended by promoting the coupling constant between the particles to a matrix that acts on the unit vectors, representing a type of generalized frustration. In a recent paper we have analyzed in detail the role of the coupling matrix for $D=2$. Here we extend this analysis to arbitrary dimensions, presenting a study of synchronous states and their stability. We show that when the natural frequencies of the particles are set to zero, the system converges either to a stationary synchronized state with well defined phase, or to an effective two-dimensional dynamics, where the synchronized particles rotate on the sphere. The stability of these states depend on the eigenvalues and eigenvectors of the coupling matrix. When the natural frequencies are not zero, synchronization depends on whether $D$ is even or odd. In even dimensions the transition to synchronization is continuous and rotating states are replaced by active states, where the order parameter rotates while its module oscillates. If $D$ is odd the phase transition is discontinuous and active states are suppressed, occurring only for a restricted class of coupling matrices.
	
\end{abstract}

\maketitle

\section{Introduction}

The Kuramoto model describes the synchronization dynamics of a set of interacting oscillators \cite{Kuramoto1975,Kuramoto1984}. The model has been used to describe several natural and artificial systems, such as circadian rhythms \cite{yamaguchi2003,bick2020understanding}, power grids \cite{filatrella2008analysis,Nishikawa_2015}, neuronal networks \cite{cumin2007generalising,bhowmik2012well,ferrari2015phase,reis2021bursting} and metronomes \cite{Pantaleone2002}.  The oscillators are represented by their phases $\theta_i$ and are coupled according to the equations 
\begin{equation}
	\dot{\theta}_i = \omega_i + \frac{k}{N} \sum_{j=1}^N \sin{(\theta_j-\theta_i)}
	\label{kuramoto}
\end{equation}
where  $\omega_i$ are their natural frequencies, selected from a symmetric distribution $g(\omega)$, $k$ is the coupling strength and $i = 1, ..., N$. The complex order parameter 
\begin{equation}
	z = p e^{i \psi} \equiv \frac{1}{N} \sum_{i=1}^N e^{i\theta_i}
	\label{paraord}
\end{equation}
measures the degree of phase synchronization of the system: disordered motion results in $p \approx 0$ and coherent motion in $p \approx 1$. Kuramoto showed that, in the limit where $N \rightarrow \infty$, the onset of synchronization could be described as a continuous phase transition, where $p$ remains very small for $0 < k < k_c =  2/\pi g(0)$ and increases as $p = \sqrt{1-k_c/k}$ for $ k > k_c$ \cite{Acebron2005,Rodrigues2016}. 

Recently, Chandra et al \cite{chandra2019continuous} have shown that Kuramoto oscillators could be reinterpreted as unit vectors $\vec{\sigma_i} = (\cos{\theta_i},\sin{\theta_i})$ rotating on the unit circle. Using Eq.(\ref{kuramoto}) it is easy to show that the dynamics of $\vec{\sigma}_i$ is given by 
\begin{equation}
	\frac{d \vec{\sigma_i}}{d t} = \mathbf{W}_i \vec{\sigma_i} + \frac{k}{N} \sum_j [\vec{\sigma_j} - (\vec{\sigma_i}\cdot \vec{\sigma_j}) \vec{\sigma_i}]
	\label{eq3}
\end{equation}
where  $\mathbf{W}_i$ is the anti-symmetric natural frequency matrix 
\begin{equation}
	\mathbf{W}_i = \left( 
	\begin{array}{cc}
		0 & -\omega_i \\
		\omega_i & 0
	\end{array}
	\right).
	\label{wmat}
\end{equation}
The complex order parameter  $z$, Eq.(\ref{paraord}), can be conveniently written in terms of the real vector
\begin{equation}
	\vec{p} = \frac{1}{N}\sum_i \vec{\sigma_i} = (p\cos\psi,p\sin\psi)
	\label{vecpar}
\end{equation}
describing the center of mass of the system. 

Eq.(\ref{eq3}) can be extended to higher dimensions by simply considering unit vectors $\vec\sigma_i$ in D-dimensions, rotating on the surface of the corresponding (D-1) unit sphere (see also \cite{Tanaka2014,Strogatz2019higher,crnkic2021synchronization}). The matrices $\mathbf{W}_i$ become $D \times D$ anti-symmetric matrices containing the $D(D-1)/2$ natural frequencies of each oscillator. It has been shown, in particular, \cite{chandra2019continuous} that the system exhibits discontinuous  phase transitions in odd dimensions, which attracted a lot of attention \cite{markdahl2020high,markdahl2021almost,dai2021d}. Previous examples of discontinuous transitions in the Kuramoto model required interactions via scale-free networks \cite{gomez2011explosive}, specific correlations between coupling strengths and natural frequencies \cite{zhang2013explosive}, or multilayer networks \cite{kachhvah2021explosive}.

In a previous paper \cite{barioni2021ott} we have further extended the multidimensional Kuramoto model by replacing the coupling constant $k$ by a coupling matrix $\mathbf{K}$, changing the equations to
\begin{equation}
	\frac{d \vec{\sigma_i}}{d t} = \mathbf{W}_i \vec{\sigma_i} + \frac{1}{N} \sum_j [{\mathbf K} \vec{\sigma_j} - (\vec{\sigma_i}\cdot {\mathbf K} \vec{\sigma_j}) \vec{\sigma_i}].
	\label{kuragen}
\end{equation}
The coupling matrix can be interpreted as a generalized frustrated model, as $\mathbf{K}$ rotates $\vec\sigma_j$ hindering its alignment with $\vec\sigma_i$ and inhibiting synchronization. Phase frustration is often associated with time-delayed couplings and can describe imperfections and heterogeneities of real oscillatory systems \cite{yue2020model}. It has been used to characterize various systems,  including Josephson junctions \cite{wiesenfeld1996synchronization}, power grids \cite{Nishikawa_2015} and seismology \cite{vasudevan2015earthquake}. Using Eq.(\ref{vecpar}) for the order parameter, Eq.(\ref{kuragen}) can also be written as
\begin{equation}
	\frac{d \vec{\sigma_i}}{d t} = \mathbf{W}_i \vec{\sigma_i} +  [ \mathbf{K} \vec{p}- (\vec{\sigma_i}\cdot \mathbf{K} \vec{p}) \vec{\sigma_i}].
	\label{kuramotogenk}
\end{equation}
Norm conservation, $|\vec{\sigma_i}|=1$, is guaranteed for any regular matrix ${\mathbf K}$, as can be seen by taking the scalar product of Eqs.(\ref{kuragen}) or (\ref{kuramotogenk}) with $\vec{\sigma_i}$. 

In \cite{buzanello2022matrix} we have analysed in detail the two-dimensional case for arbitrary matrices $\mathbf{K}$. Because $\mathbf{K}$ and $\mathbf{W}_i$ do not generally commute, the average value of the natural frequencies, $\omega_0$, plays an important role in the dynamics. We used the Ott-Antonsen ansatz \cite{Ott2008} to show that the type of synchronous state achieved by the oscillators depend on whether the eigenvalues of $\mathbf{K}$ are real or complex and on $\omega_0$. If the eigenvectors are real and at least one of the eigenvalues is larger than $\lambda_c$, then if $\omega_0=0$ the oscillators converge to the location on the circle indicated by the corresponding eigenvector, breaking the rotational symmetry. For complex eigenvectors and $\omega_0=0$ the synchronized oscillators rotate with angular velocity proportional to the imaginary part of the eigenvalue and the module of the order parameter oscillates in time, generating {\it active states} where the cluster of synchronized oscillators expand and contract as they rotate. For specific choices of $\mathbf{K}$ the oscillators can rotate while the module of $\vec{p}$ remains constant, corresponding to the Kuramoto-Sakaguchi model \cite{sakaguchi1986soluble}.

Here we use a different approach to extend the analysis to arbitrary dimensions. We will show that the real and complex eigenvalues and eigenvectors of $\mathbf{K}$ still play a key role in determining the character and stability of the synchronized state. All analytical results will be based on a simplified version of Eq. (\ref{kuragen}), taking the matrix of natural frequencies as zero:
\begin{equation}
	\frac{d \vec{\sigma_i}}{d t} = \frac{1}{N} \sum_j [{\mathbf K} \vec{\sigma_j} - (\vec{\sigma_i}\cdot {\mathbf K} \vec{\sigma_j}) \vec{\sigma_i}].
	\label{kuragenw}
\end{equation}

\noindent{\it {\bf Outline and summary of results:}}

The eigenvalues of the real matrix $\mathbf{K}$ are either all real or appear in pairs of complex conjugate. Odd dimensional matrices must have at least one real eigenvalue but for even dimensions there might be $D$ real eigenvalues, $D/2$ pairs of complex eigenvalues or combinations of the two. In the next sections we consider these cases separately, as they need slightly different considerations. We define
\begin{equation}
	\lambda_M(\mathbf{K}) = \mbox{max} \{ \mbox{ Re }( \lambda) \, \, | \, \lambda \, \mbox{ is an eigenvalue of } \,\mathbf{K} \}.
\end{equation}

We shall first derive analytical results for Eq.(\ref{kuragenw}). We will show that if $\vec{v}$ is a real eigenvector of $\mathbf{K}$ with eigenvalue $\lambda$, then $\sigma_i = \vec{v}$ is always a solution of Eq.(\ref{kuragenw}). The solution is stable if $\lambda = \lambda_M (\mathbf{K}) >0$, i.e., $\lambda$ is positive and larger than all other real eigenvalues of $\mathbf{K}$ (section II) and, if there are complex eigenvalues, it has also to be larger than the real part of all such eigenvalues (section III). 

If a complex eigenvector $\vec{u} = \vec{v}_1 + i\, \vec{v}_2$ with eigenvalue $\lambda = \lambda_1 + i\, \lambda_2$ exists such that $\lambda_1 = \lambda_M (\mathbf{K}) > 0$, then the stable solution of Eq.(\ref{kuragenw}) is of the form $\sigma_i = \alpha(t) \vec{v}_1 + \beta(t) \vec{v}_2$ where $\alpha$ and $\beta$ can be calculated explicitly, and are periodic functions of time with period $2\pi/\lambda_2$. This is a 'rotating solution', as the order parameter rotates in the plane defined by  $\vec{v}_1$ and $\vec{v}_2$ (section IV). We note that these analytical solutions of Eq.(\ref{kuragenw}),  $\sigma_i = \vec{v}$ or $\sigma_i = \alpha(t) \vec{v}_1 + \beta(t) \vec{v}_2$,  correspond to perfect synchronization, as all oscillators have identical behavior.

In section V we reintroduce the matrices $\mathbf{W}_i$ and consider qualitatively the effects of non-zero natural frequencies.  When natural frequencies of each oscillator are drawn from a distribution as in Eq. (\ref{kuramotogenk}), synchronization is generally partial and it matters if $D$ is even or odd. For odd dimensions partial sync requires only $\lambda_M > 0$ and the transition to synchronization is discontinuous \cite{chandra2019continuous}. If there are multiple real eigenvalues, the eigenvector $\vec{v}$ with largest positive eigenvalue defines an approximate stable solution where $\vec{\sigma}_i = \vec{v}$. It is not an exact solution because synchronization is only partial. If $\lambda_M$ corresponds to a complex eigenvalue, then the rotating solution is stable only if all other real eigenvalues are negative. This means that the rotation solution is suppressed by the real eigenvectors with positive eigenvalues even if the complex eigenvalue satisfies $\lambda_1 = \lambda_M$. For even dimensions synchronization requires $\lambda_M > \lambda_c$, where $\lambda_c > 0$ depends on $D$ \cite{chandra2019continuous,barioni2021ott} and the phase transition is continuous. In this case the dynamics is determined exclusively by the eigenvector with $\lambda=\lambda_M$. Finally we distinguish between 'pure rotations', when the cluster of synchronized oscillators rotate but the module of the order parameter remains contant, and 'active states', where the module of the order parameter oscillates as it rotates. The first case occurs only if $\vec{v}_1 \cdot \vec{v}_2 = 0$ whereas the second is more general and occurs whenever $\vec{v}_1 \cdot \vec{v}_2 \neq 0$.

In section VI we illustrate our findings with numerical simulations and summarized our conclusions in section VII.

\section{Real eigenvalues: static solutions}

In this section we assume that all eigenvalues of $\mathbf{K}$ are real. Let
\begin{equation}
	\mathbf{K} \vec{v}_\gamma = \lambda_\gamma \vec{v}_\gamma 
\end{equation}
with $\gamma=1,2,\dots,D$ and $|\vec{v}_\gamma|=1$. If the eigenvalues are non-degenerated, there are $D$ synchronized solutions of Eq. (\ref{kuragenw}), given by $\hat{\sigma}_i = \vec{v}_\gamma$. This can immediately verified by noticing that $(\vec{v}_\gamma\cdot {\mathbf K} \vec{v}_\gamma) \vec{v}_\gamma = {\mathbf K} \vec{v}_\gamma = \lambda_\gamma \vec{v}_\gamma$. Therefore, the eigenvectors of $\mathbf{K}$ indicate special positions on the sphere where all the oscillators stay in equilibrium, corresponding to static solutions.

Next we show that the solution $\hat{\sigma}_i = \vec{v}_\alpha$ is stable if $\lambda_\alpha > 0$ and $\lambda_\alpha > \lambda_\gamma$ for all $\gamma \neq \alpha$. Let
\begin{equation}
	\hat{\sigma}_i = \vec{v}_\alpha + \vec{x}_i
	\label{xreal}
\end{equation}
where $\vec{x}_i$ is a small perturbation. Norm conservation requires $\vec{v}_\alpha \cdot \vec{x}_i = 0$. Substituting (\ref{xreal}) in (\ref{kuragenw}) and discarding terms in $|\vec{x}_i|^2$ leads to
\begin{equation}
	\dot{\vec{x}}_i = -\lambda_\alpha \vec{x}_i + \frac{1}{N} \sum_{j=1}^N \left[\mathbf{K} \vec{x}_j
	- (\vec{v}_\alpha \cdot \mathbf{K} \vec{x}_j) \vec{v}_\alpha \right].
	\label{eqxreal}
\end{equation}
In order to evaluate $\mathbf{K} \vec{x}_j$ we expand the perturbations in the basis of eigenvectors of $\mathbf{K}$, taking into account that these vectors are not generally orthogonal, as $\mathbf{K}$ is not necessarily symmetric:
\begin{equation}
	\vec{x}_j = \sum_\gamma a_{j\gamma} \vec{v}_\gamma.
	\label{xex}
\end{equation}
Imposing $\vec{v}_\alpha \cdot \vec{x}_j = 0$ leads to
\begin{equation}
	a_{j\alpha} = - \sum_{\gamma \neq \alpha} g_{\alpha \gamma} a_{j\gamma}
\end{equation}
where we have defined
\begin{equation}
	g_{\gamma \beta} = g_{\beta \gamma} \equiv \vec{v}_\gamma \cdot \vec{v}_\beta
\end{equation}
with $g_{\beta \beta} =1$. This allows us to rewrite Eq.(\ref{xex}) as
\begin{equation}
	\vec{x}_j = \sum_{\gamma \neq \alpha} a_{j\gamma} \vec{V}_\gamma.
	\label{xex2}
\end{equation}
where
\begin{equation}
	\vec{V}_\gamma = \vec{v}_\gamma - g_{\alpha \gamma} \vec{v}_\alpha.
	\label{valpha}
\end{equation}
We can now compute
\begin{equation}
	\mathbf{K} \vec{V}_\gamma = \lambda_\gamma \vec{v}_\gamma - \lambda_\alpha g_{\alpha \gamma} \vec{v}_\alpha;
\end{equation}
\begin{equation}
	(\vec{v}_\alpha \cdot\mathbf{K} \vec{V}_\gamma) \vec{v}_\alpha = \lambda_\gamma g_{\alpha \gamma} \vec{v}_\alpha - \lambda_\alpha g_{\alpha \gamma} \vec{v}_\alpha;
\end{equation}
\begin{equation}
	\mathbf{K} \vec{V}_\gamma - (\vec{v}_\alpha \cdot\mathbf{K} \vec{V}_\gamma) \vec{v}_\alpha = \lambda_\gamma \vec{V}_\gamma. 
	\label{kvalpha}
\end{equation}
Substituting Eq.(\ref{xex2}) into (\ref{eqxreal}) and using (\ref{valpha}) and (\ref{kvalpha}) we obtain equations for the coefficients $a_{i\gamma}$:
\begin{equation}
	\dot{a}_{i\gamma} = -\lambda_\alpha a_{i\gamma} + \frac{1}{N}\sum_j \lambda_\gamma a_{j\gamma} \equiv \sum_j M_{ij} a_{j\gamma}.
	\label{linearsys}
\end{equation}
The eigenvalues of the tangent matrix $M$ are $-\lambda_\alpha$ (with degeneracy $(N-1)$) and $-\lambda_\alpha + \lambda_\gamma$. Therefore, solution $\hat{\sigma}_i = \vec{v}_\alpha$ is a stable node if $\lambda_\alpha = \lambda_M > 0$, i.e., $\vec{v}_\alpha$ is the eigenvector of $\mathbf{K}$ with the largest eigenvalue and $\lambda_\alpha > 0$.

\section{Complex eigenvalues: static solutions}

The previous calculation assumed that all eigenvalues of $\mathbf{K}$ were real. We now consider the case where $\lambda_\alpha$ and $\vec{v}_\alpha$ are real but $\mathbf{K}$ has a pair of complex eigenvectors
\begin{equation}
	\mathbf{K} \vec{u} = \lambda \vec{u}; \qquad \mathbf{K} \vec{u}^{\,*} = \lambda^* \vec{u}^{\, *}.
\end{equation}
We first compute the stability of the solution $\hat{\sigma}_i = \vec{v}_\alpha$ in this situation. Writing 
\begin{equation}
	\vec{u} = \vec{v}_1 + i \vec{v}_2; \qquad \lambda = \lambda_1 + i \lambda_2
	\label{v1v2}
\end{equation}
we can expand the perturbations $\vec{x}_i$ in terms of the vectors $\vec{v}_1$, $\vec{v}_2$ and the remaining $D-2$ real vectors $\vec{v}_\gamma$. Using the orthogonality between $\vec{v}_\alpha$ and $\vec{x}_i$, and Eq.(\ref{valpha}), we can still write
\begin{equation}
	\vec{x}_j = \sum_{\gamma \neq \alpha} a_{j\gamma} \vec{V}_\gamma.
	\label{xex22}
\end{equation}
where $\vec{V}_\gamma$ for $\gamma=1$ and $\gamma=2$ involve the vectors defined in Eq.(\ref{v1v2}), which are not themselves eigenvectors of $\mathbf{K}$. Instead, they satisfy the equations
\begin{eqnarray}
	\mathbf{K} \vec{v}_1 & = & \lambda_1 \vec{v}_1 - \lambda_2 \vec{v}_2 \\
	\mathbf{K} \vec{v}_2 & = & \lambda_1 \vec{v}_2 + \lambda_2 \vec{v}_1
	\label{kv1v2}
\end{eqnarray}
which lead to
\begin{eqnarray}
	\mathbf{K} \vec{V}_1 - (\vec{v}_\alpha \cdot \mathbf{K} \vec{V}_1) \vec{v}_\alpha & = & \lambda_1 \vec{V}_1 - \lambda_2 \vec{V}_2 \\
	\mathbf{K} \vec{V}_2 - (\vec{v}_\alpha \cdot \mathbf{K} \vec{V}_2) \vec{v}_\alpha & = & \lambda_1 \vec{V}_2 + \lambda_2 \vec{V}_1.
\end{eqnarray}
For $\gamma \neq 1,2, \alpha$ Eqs. (\ref{kvalpha}) and (\ref{linearsys}) remain valid. For $\gamma=1,2$ we obtain
\begin{eqnarray}
	\dot{a}_{i 1} &=& -\lambda_\alpha a_{i1} + \frac{1}{N}\sum_j (\lambda_1 a_{j1} + \lambda_2 a_{j2}) \\
	\dot{a}_{i 2} &=& -\lambda_\alpha a_{i2} + \frac{1}{N}\sum_j (\lambda_1 a_{j2} - \lambda_2 a_{j21})
\end{eqnarray}
representing a linear system of dimension $2N \times 2N$. The corresponding eigenvalues are $-\lambda_\alpha$, $(2N-2)$-degenerated, and $(-\lambda_\alpha + \lambda_1) \pm i \lambda_2$. Therefore, the solution $\hat{\sigma}_i = \vec{v}_\alpha$ can be characterized as a stable focus if $\lambda_\alpha >0 $,  $\lambda_\alpha > Re(\lambda_\gamma)$ and $\lambda_\alpha$ is larger than all other real eigenvalues of $\mathbf{K}$. In other words $\lambda_\alpha = \lambda_M >0$. The spiral behavior of the perturbation occurs only in the $\vec{v}_1-\vec{v}_2$ plane and with frequency $\lambda_2$. In the other directions the solution behaves as a stable node. The analysis extends directly to the case of more then one pair of complex eigenvectors.

\section{Complex eigenvalues: rotating solutions}

\subsection{Rotating solutions}

We finally consider the case where the dynamics is dictated by a pair of complex conjugate eigenvectors. We search for solutions of Eq.(\ref{kuragenw}) of the form
\begin{equation}
	\hat\sigma_i  \equiv \vec{w} = \alpha \vec{v}_1 + \beta \vec{v}_2 
	\label{complexsol}
\end{equation}
where $\vec{v}_1$ and $\vec{v}_2$ are the real and imaginary parts of the complex eigenvector $\vec{u}$ as defined by Eqs.(\ref{v1v2}). Normalization requires
\begin{equation}
	\alpha^2 + \beta^2 + 2\alpha\beta g_{12} = 1.
\end{equation}

Substituting (\ref{complexsol}) into (\ref{kuragenw}) and using (\ref{kv1v2}) we find that $\alpha$ and $\beta$ must satisfy the equations
\begin{eqnarray}
	\dot{\alpha} = \lambda_2 [\beta + \alpha g_{12} (\alpha^2 - \beta^2)] \\ 
	\dot{\beta} = \lambda_2 [-\alpha + \beta g_{12} (\alpha^2 - \beta^2)].
	\label{abdot1}
\end{eqnarray}
Defining $A=(\alpha+\beta)/\sqrt{2}$ and $B=(\alpha-\beta)/\sqrt{2}$ these equations simplify to
\begin{eqnarray}
	\dot{A} = -\lambda_2 B [1 - 2 g_{12} A^2] \\ 
	\dot{B} = \lambda_2 A [1 + 2 g_{12} B^2]
	\label{eqsab}
\end{eqnarray}
and the normalization condition becomes $A^2 (1+g_{12}) + B^2(1-g_{12}) = 1$. These equations can be solved analytically (see appendix \ref{appa}) and we find:
\begin{eqnarray}
	A &=& \frac{\cos(\lambda_2 t)}{\sqrt{1+ g_{12} \cos(2\lambda_2 t)}} \\
	B &=& \frac{\sin(\lambda_2 t)}{\sqrt{1+ g_{12} \cos(2\lambda_2 t)}},
	\label{absol}
\end{eqnarray}
which are periodic oscillations with period $2\pi/\lambda_2$. It can be checked that normalization is preserved at all times, i.e., $\alpha \dot{\alpha} + \beta \dot{\beta} + g_{12}(\alpha \dot{\alpha}+ \beta \dot{\beta}) = 0$. This solution generalizes the Kuramoto-Sakaguchi model where rotations of the order parameter in 2D are induced by frustration \cite{sakaguchi1986soluble}.

It is convenient to introduce a new unit vector $\vec{z}$ perpendicular to $\vec{w}$ as:
\begin{equation}
	\vec{z} = \eta [ (\beta + \alpha g_{12}) \vec{v}_1 - (\alpha + \beta g_{12}) \vec{v}_2]
\end{equation}
where $\eta = (1-g_{12}^2)^{-1/2}$ ensures the normalization. It can be shown that
\begin{equation}
	\dot{\vec{w}} = \lambda_2 \eta^{-1} (\alpha^2 + \beta^2) \vec{z}
	\label{wdot}
\end{equation}
and
\begin{equation}
	\dot{\vec{z}} = - \lambda_2 \eta^{-1} (\alpha^2 + \beta^2) \vec{w}
	\label{zdot}
\end{equation}
which would characterize a harmonic oscillator if $\alpha$ and $\beta$ where not time dependent. From $\vec{z} \cdot \vec{w} = 0$ it follows that $\dot{\vec{w}} \cdot \vec{w} = 0$ showing again that the norm of $\vec{w}$ is preserved by the dynamics.

\subsection{Stability}

With the rotating state $\vec{w}$ fully characterized we can now analyse its stability. As before we write
\begin{equation}
	\hat{\sigma}_i = \vec{w} + \vec{x}_i
	\label{xcpx}
\end{equation}
and expand the perturbation as
\begin{equation}
	\vec{x}_i = a_{iw} \vec{w} + a_{iz} \vec{z} + \sum_\gamma a_{i\gamma} \vec{v}_\gamma = a_{iz} \vec{z} + \sum_\gamma a_{i\gamma} \vec{V}_\gamma
	\label{xxcpx}
\end{equation}
where $\gamma$ runs over the real eigenvectors of $\mathbf{K}$. In the last equality we have used $\vec{w} \cdot \vec{x}_i = 0$ and $\vec{V}_\gamma = \vec{v}_\gamma - g_{w\gamma} \vec{w}$. Substituting (\ref{xcpx}) into Eq.(\ref{kuragenw}) and linearizing we obtain
\begin{equation}
	\dot{\vec{x}}_i = -(\vec{x}_i \cdot \mathbf{K }\vec{w})\vec{w} - (\vec{w} \cdot \mathbf{K} \vec{w}) \vec{x}_i + \frac{1}{N} \sum_{j=1}^N \left[\mathbf{K} \vec{x}_j
	- (\vec{w} \cdot \mathbf{K} \vec{x}_j) \vec{w} \right].
	\label{eqxcpx}
\end{equation}
Finally, using Eq.(\ref{xxcpx}) we obtain the equations describing the dynamics of the coefficients. The details are in the Appendix \ref{appb}. We first consider the equation for $a_{i\gamma}$:
\begin{eqnarray}
	\dot{a}_{i\gamma} &=& - a_{i\gamma} [\lambda_1 - \lambda_2 g_{12}(\alpha^2-\beta^2)] + \frac{1}{N} \sum_j a_{j\gamma} \lambda_\gamma.
	\label{eqgamma}
\end{eqnarray}
For $g_{12}=0$ this is equivalent to the linear system in Eq.(\ref{linearsys}) and the condition for $a_{i\gamma}$ converge to zero is $\lambda_1 > \lambda_\gamma$. For $g_{12}\neq 0$ the time-dependent factors $\alpha$ and $\beta$ can be eliminated with the transformation
\begin{equation}
	a_{iz} = A_{iz} \exp{\left\{ - \lambda_2 g_{12} \int_0^t [\alpha^2(t')-\beta^2(t') ]dt' \right\} }
\end{equation}
resulting in
\begin{eqnarray}
	\dot{A}_{i\gamma} &=& - A_{i\gamma} \lambda_1  + \frac{1}{N} \sum_j A_{j\gamma} \lambda_\gamma.
\end{eqnarray}
Since $\alpha^2(t) - \beta^2(t) = 2 A(t) B(t)$, its integral over one period is zero and the integral over any time interval is bounded. Therefore, the condition $\lambda_1 > \lambda_\gamma$ suffices for the coefficients $a_{i\gamma}$ to go to zero.

The equation for $a_{iz}$ is more complicated. However, assuming that $a_{i\gamma}$ converge to zero, they simplify to
\begin{eqnarray}
	\dot{a}_{iz} &=& - a_{iz} [\lambda_1 - \lambda_2 g_{12}(\alpha^2-\beta^2)] + \frac{1}{N} \sum_j a_{jz} [\lambda_1 + \lambda_2 g_{12}(\alpha^2-\beta^2)].
	\label{eqzdot}
\end{eqnarray}
Using again the fact that the time dependent factors involving $\alpha^2 - \beta^2$ have bounded time integrals, it suffices that $\lambda_1 > 0$ to guarantee the that the coefficients $a_{iz}$ also go to zero (see Appendix \ref{appb}). The stability condition for the rotating state is, therefore, that the real part of the complex eigenvalue $\lambda$ is positive and larger than all real eigenvalues of $\mathbf{K}$.

\section{Non-zero natural frequencies}

\subsection{The role of dimensionality}

The analysis presented in the last sections were possible because the matrices of natural frequencies $\mathbf{W}_i$ were set to zero. When these are turned on, the effects of dimensionality immediately kick in. If $\mathbf{K} = k \mathbf{1}$ (scalar coupling), it has been shown that for $D$ odd the transition to synchrony is discontinuous and only requires $k>0$ \cite{chandra2019continuous}. This is related to the fact that a $D\times D$ anti-symmetric matrix $\mathbf{W}_i$ always has an eigenvector $\vec{q}_i$ with zero eigenvalue if $D$ is odd. In this case, Eq.(\ref{kuramotogenk}) has two simple stationary solutions in the limit $k \rightarrow 0^+$, given by $\vec{\sigma}_i = \pm \vec{q}_i$. However, only the solution satisfying $\vec{\sigma}_i \cdot \vec{p} > 0$ is stable. This implies that all oscillators immediately move to the hemisphere containing the vector $\vec{p}$, giving rise to partial synchronization with $p=1/2$ for $D=3$ \cite{chandra2019continuous}. 

This analysis remains true in the limit where all the eigenvalues of K are small, so that
synchronization remains discontinuous for the case of matrix coupling, requiring only one
eigenvalue to have positive real part. When the natural frequencies go to zero, we have shown that the dynamics is dominated by the eigenvector with $\lambda_M$. If $\lambda_M > 0$ corresponds to a real eigenvector $\vec{v}$, it defines the direction of the order parameter and, consequently, the hemisphere of stable solutions as discussed above. The effect of non-zero natural frequencies, therefore, is to lead to partial (as opposed to perfect) synchronization around this direction, which is confirmed by numerical simulations (see section VI).

However, if $\lambda_M$ corresponds to a complex eigenvector, the order parameter for zero natural frequencies converges to $\vec{p}=\vec{w}(t)$, which rotates. In this case numerical simulations show that for non-zero natural frequencies the oscillators follow $\vec{w}(t)$ for a while, but eventually stop rotating and converge to the direction of the real eigenvector $\vec{v}_r$ with the largest eigenvalue $\lambda_r$. The asymptotic value of the order parameter $p$ around the real eigenvector is larger than the transient value around $\vec{w}(t)$, even though $\lambda_M > \lambda_r$ (see Fig.1 (c) and (d)). We don't have an analytical understanding of this phenomenon yet, but a qualitative explanation is as follows: in the limit of small $\mathbf{K}$ and small $\mathbf{W}_i$ (in the sense of small eigenvalues), the hemisphere defining the stable solutions $\vec{\sigma}_i$  rotates, causing the oscillators to flip from $+\vec{q}_i$ to $-\vec{q}_i$ as $\vec{q}_i \cdot \vec{w}(t)$ changes sign. This constant flipping makes the oscillators fall out of sync around $\vec{w}$ and move towards the real eigenvector $\vec{v}_r$ with the largest eigenvalue $\lambda_r$ if $\lambda_r >0$. Clearly this behavior is characteristic of odd dimensions only, due to the null eigenvectors $\vec{q}_i$ of $\mathbf{W}_i$. 

For even dimensions the transition for $\mathbf{K} = k \mathbf{1}$ is continuous and requires $k > k_c$, where $k_c$ depends on the distribution of frequencies (and on $D$), similar to the usual Kuramoto model with $D=2$. We hypothesize that the same happens for the case of matrix coupling and that the distribution of natural frequencies does not change the behavior qualitatively: the equilibrium is always dominated by the eigenvector with largest eigenvalue $\lambda$ (or largest real part if the eigenvalue is complex)  and (partial) synchronization requires $\lambda > k_c$. If the eigenvector is complex the group os synchronized oscillators rotate in the plane defined by the real and imaginary parts of the corresponding eigenvector. If the eigenvector is real, the system converges to a static distribution centered on its direction. These conjectures are confirmed by numerical simulations shown in the next section.

\subsection{Active states}

An important feature of dynamics induced by the matrix coupling is the appearance of active states, where the order parameter not only rotates but its module also oscillates in time. For $D=2$ it was shown in \cite{buzanello2022matrix} that, if the eigenvalues of $\mathbf{K}$ were complex, active states would appear. The case of pure rotation (constant module) occurred only if the eigenvalues are of the form $e^{\pm i \phi}$. In this case the complex eigenvector  $\vec{u} = \vec{v}_1 + i \, \vec{v}_2$ satisfies $\vec{v}_1 \cdot \vec{v}_2 = 0$ and the system becomes identical to the frustrated Kuramoto-Sakaguchi model \cite{sakaguchi1986soluble}. In our analysis we assumed the natural frequencies to be zero, leading to full synchronization and masquerading any oscillations of the module of $p$. For non-zero natural frequencies, synchrony will be partial and we expect to see active states, unless $\vec{v}_1$ and $\vec{v}_2$ are orthogonal. This expectation is confirmed by numerical simulations.

\section{Simulations}

\begin{figure}
	\includegraphics[scale=0.20]{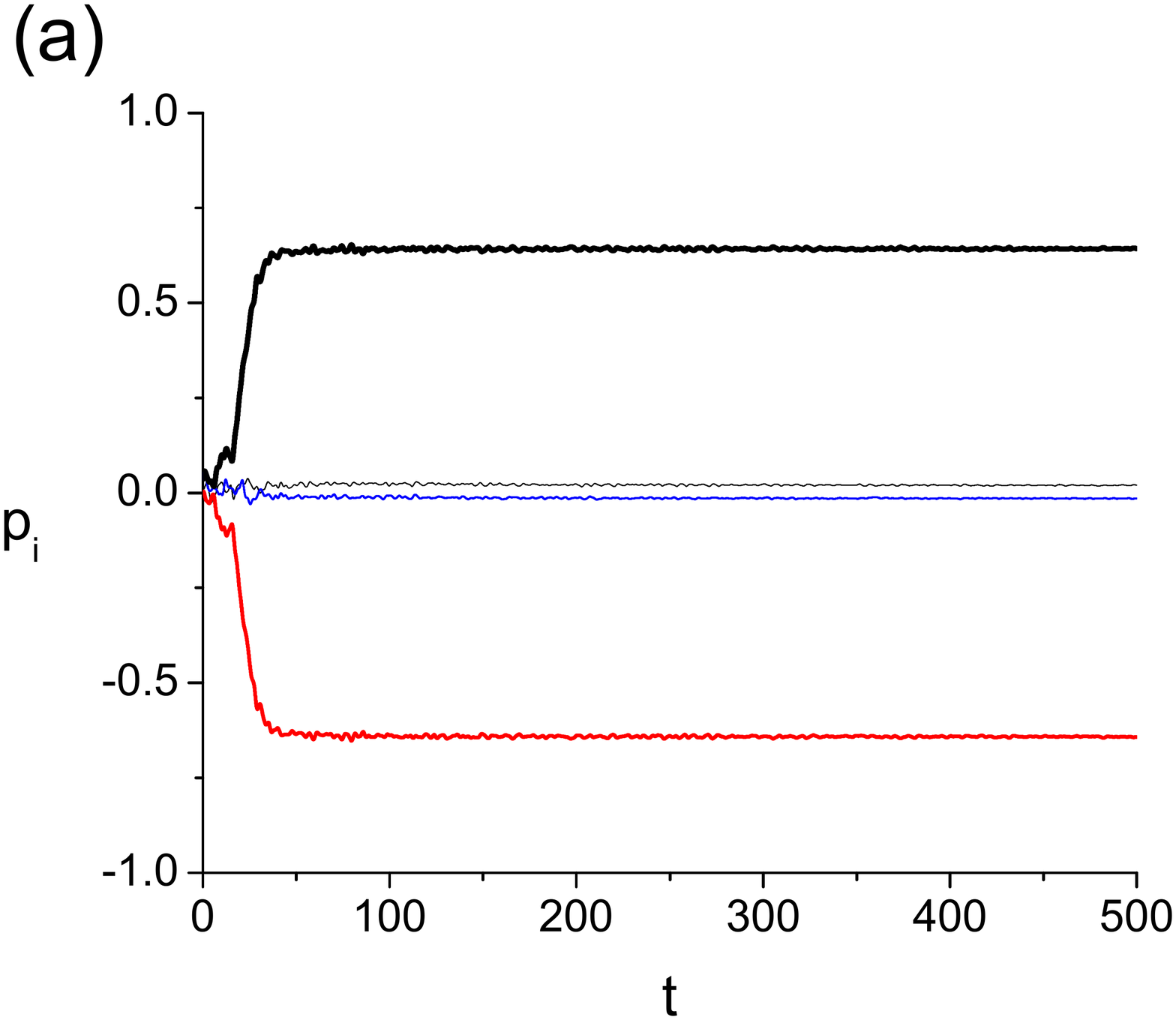} 
	\includegraphics[scale=0.20]{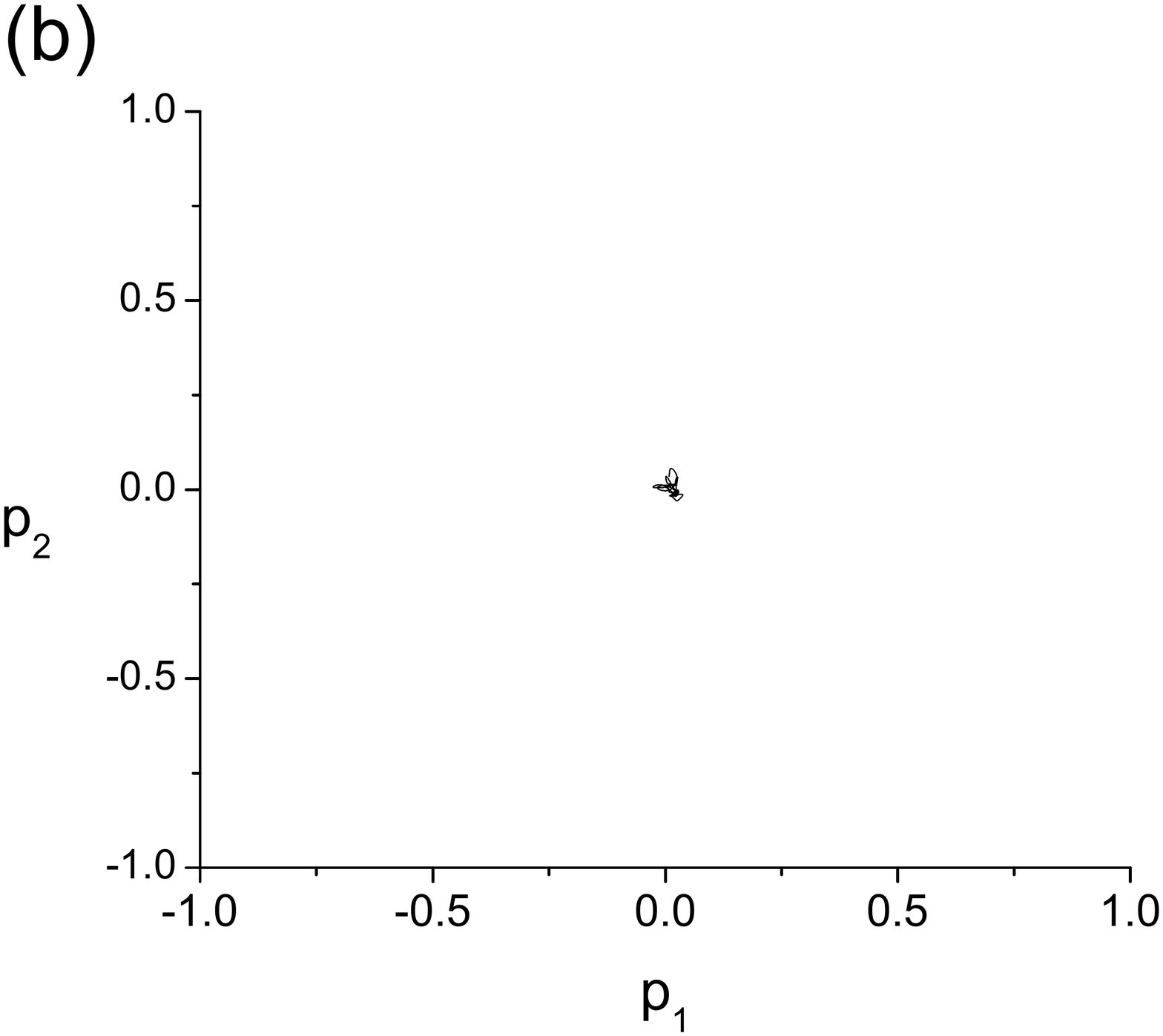} 
	\includegraphics[scale=0.20]{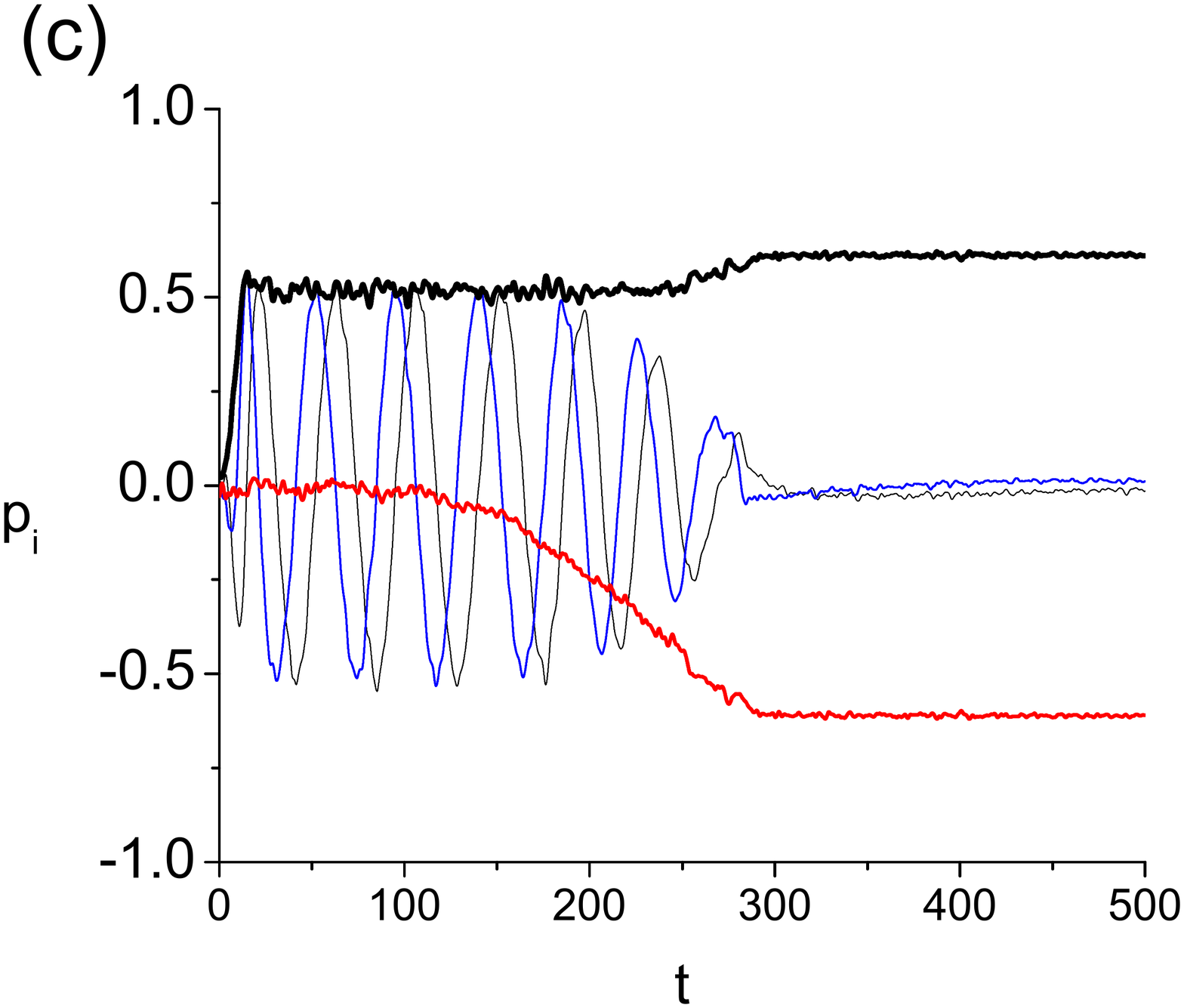} 
	\includegraphics[scale=0.20]{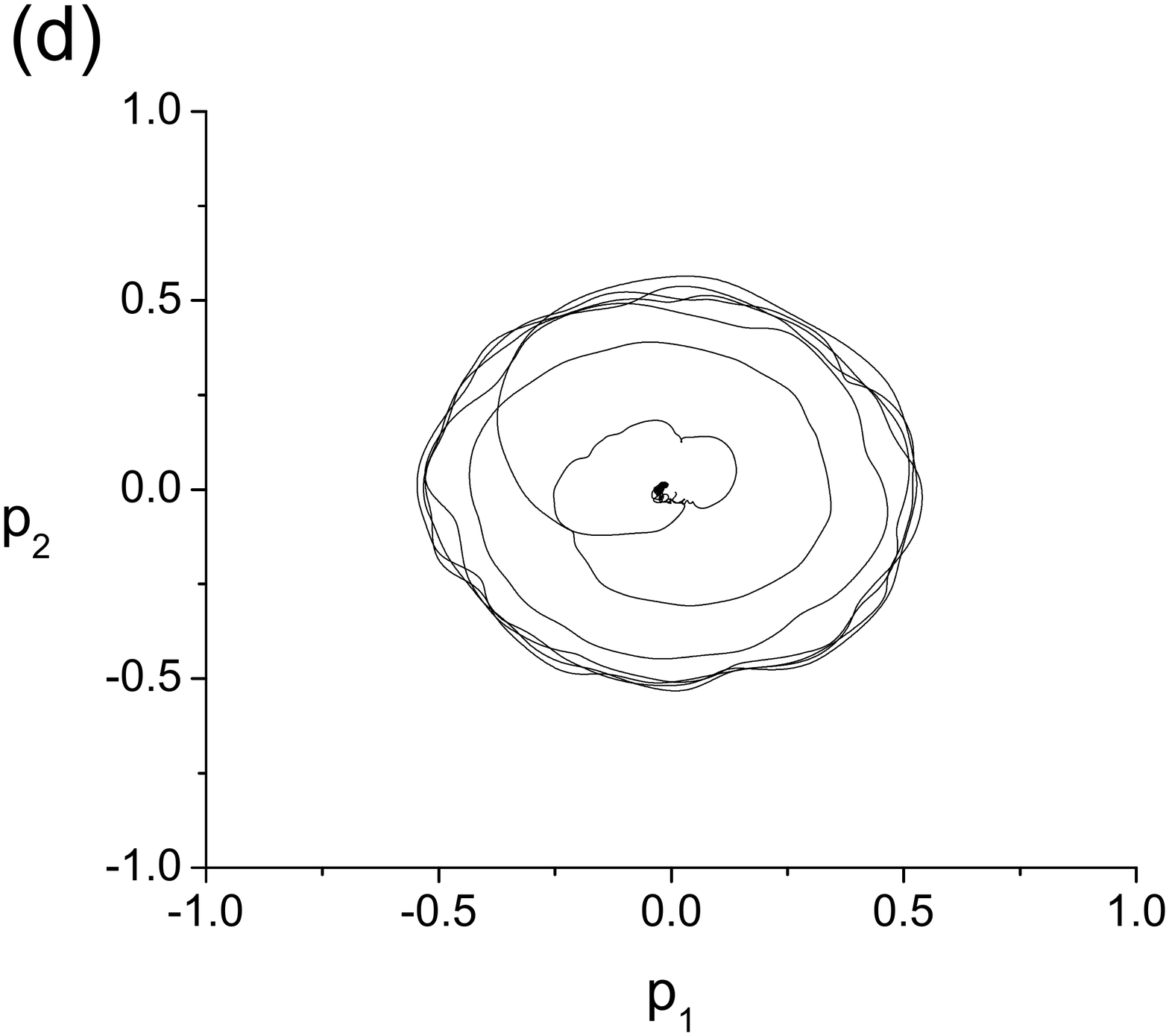} 
	\includegraphics[scale=0.20]{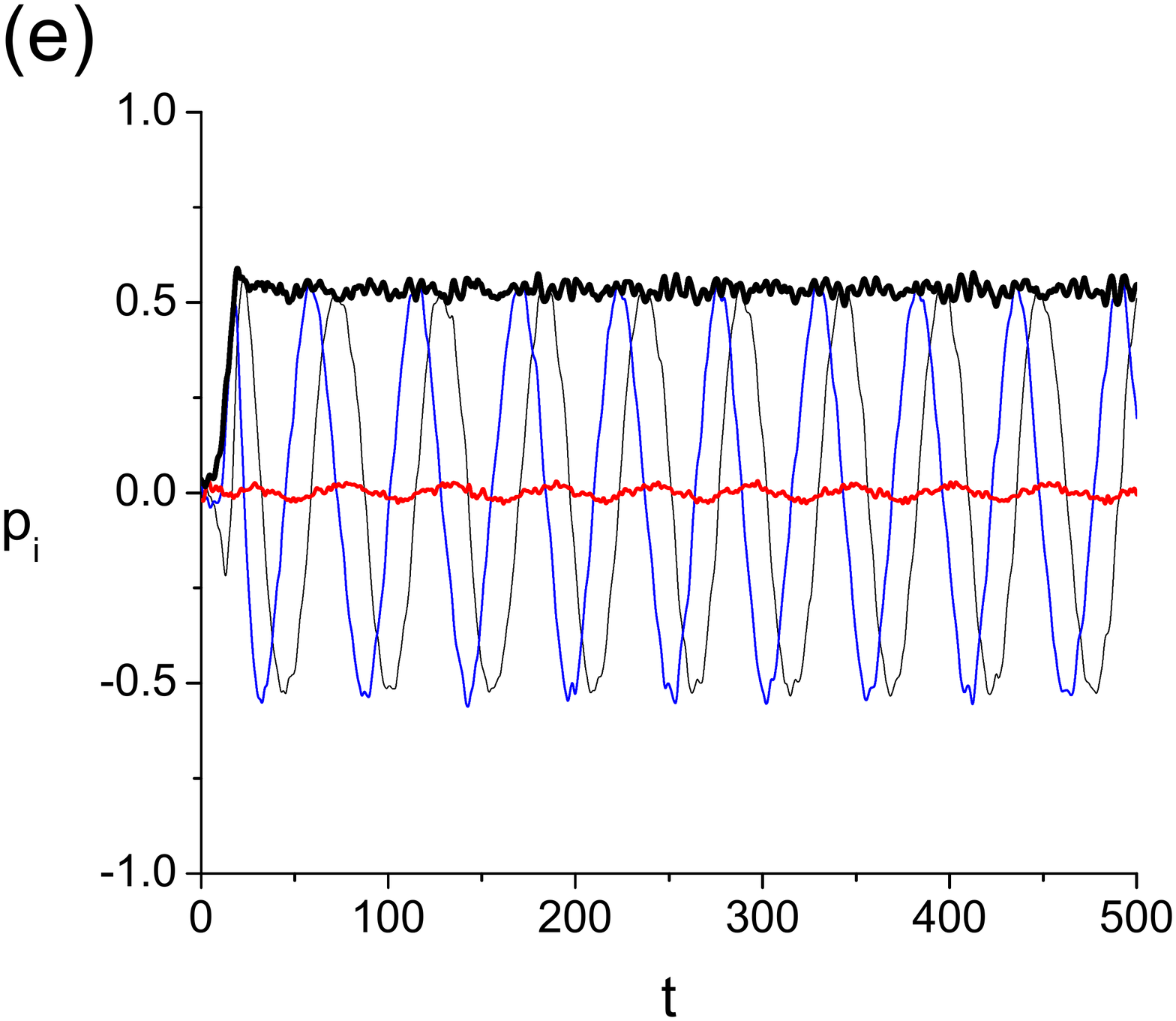} 
	\includegraphics[scale=0.20]{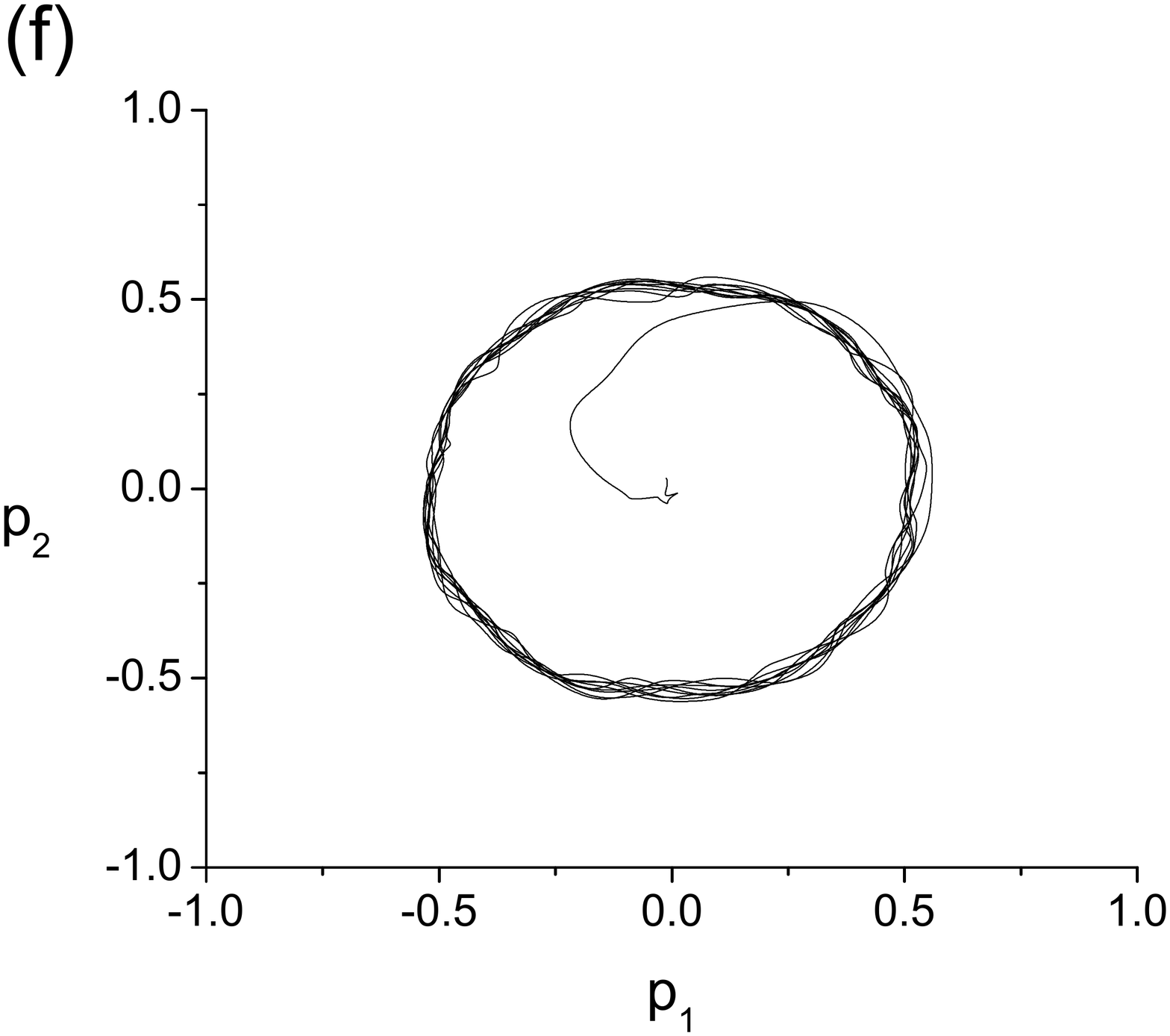} 
	\includegraphics[scale=0.20]{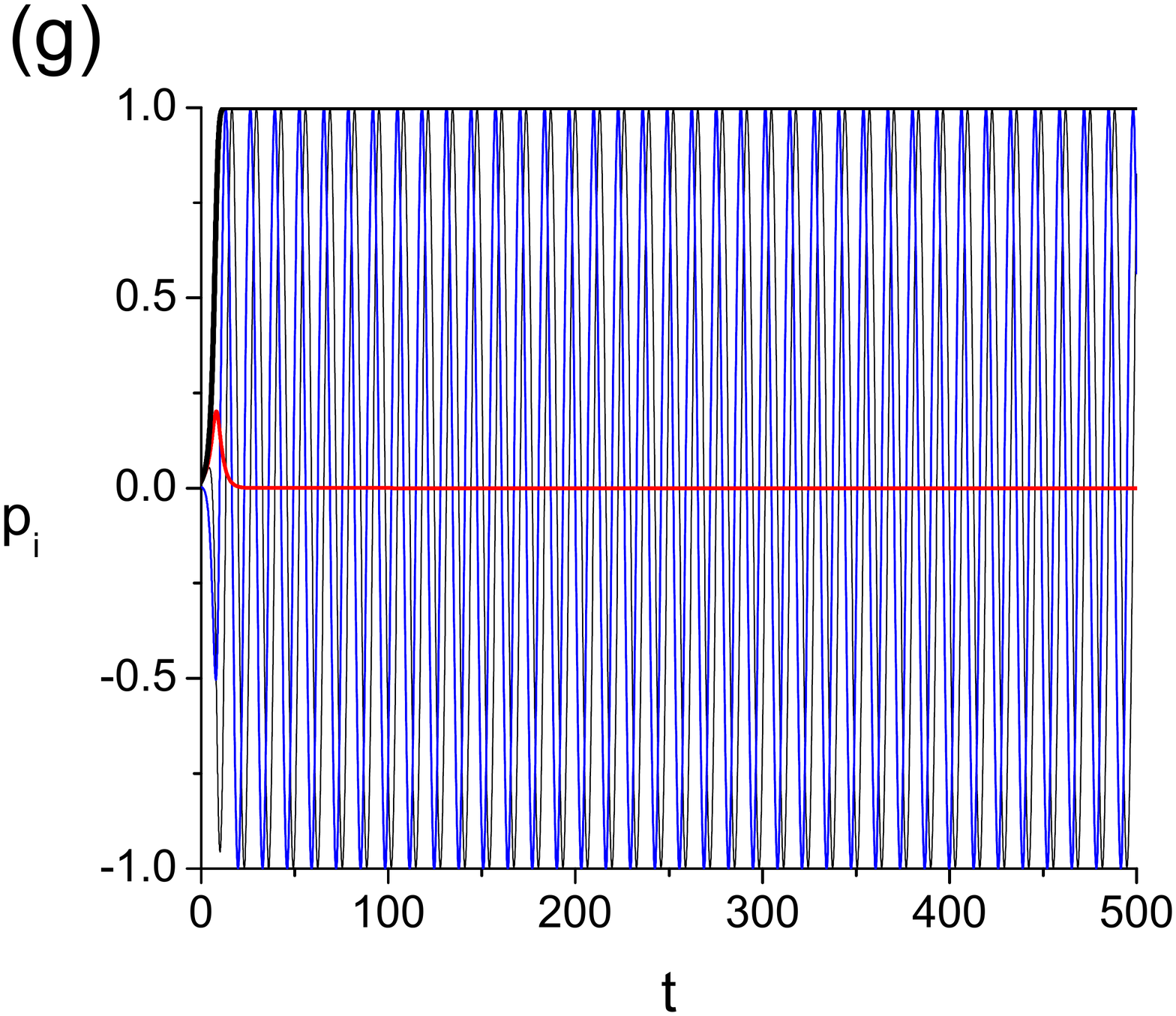} 
	\includegraphics[scale=0.20]{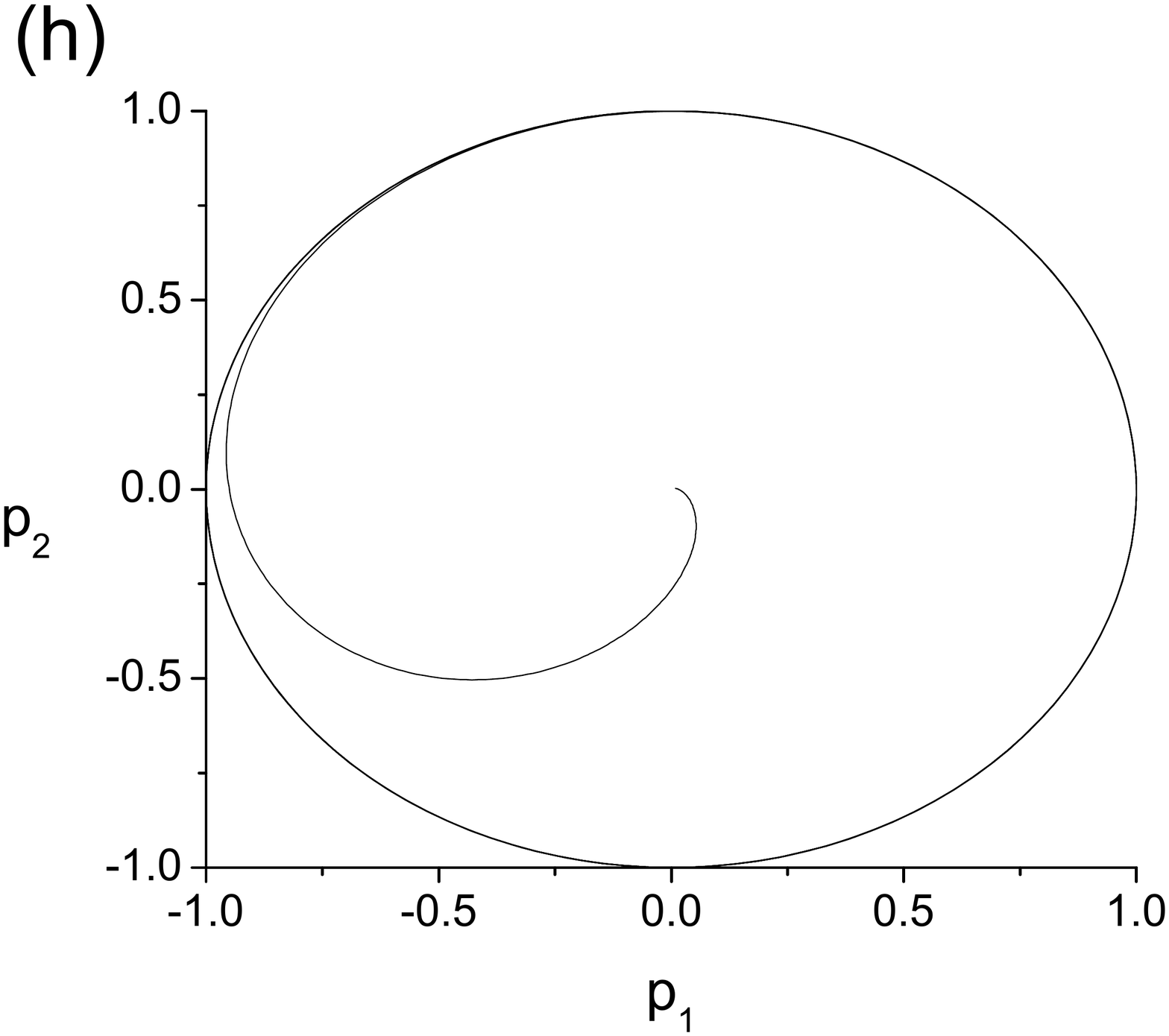} 
	\caption{3D Kuramoto model with matrix coupling as in Eq.(\ref{3d}). Left panels show the 
		time evolution of the components $p_1$ (thin black line), $p_2$ (thin blue line), $p_3$ (thick red line) and module $p$ (thick black line) of the order parameter. Right panels shown the dynamics in the $p_1 \times p_2$ plane. In panels (a) to (f) natural frequencies were sampled from a Gaussian distribution of unit width centered at zero. In panels (g) and (h) all natural frequencies were set to zero. Parameter values for the coupling matrix are: (a)-(b) $a=0.1$, $b=0.5$; (c)-(d) $a=1.0$, $b=0.5$; (e)-(f) $a=1.0$, $b=-0.5$; (g)-(h) $a=1.0$, $b=0.5$.} 
	\label{fig1}
\end{figure}

We illustrate the results derived in the previous sections with simulations in $D=3$ and $D=4$ dimensions. In all examples we use $N=1000$ oscillators. For $D=3$ we set
\begin{equation}
	\mathbf{K} = \left(
	\begin{array}{ccc}
		a \cos \alpha & a \sin \alpha & 0 \\
		-a \sin \alpha & a \cos \alpha & 0 \\
		0 & 0 & b
	\end{array}
	\right)
	\label{3d}
\end{equation}
so that the the real and complex eigenvectors are easy to identify. The eigenvalues are $a e^{\pm i\alpha}$ and $b$. The matrices of natural frequencies are given by
\begin{equation}
	\mathbf{W}_i = \left( 
	\begin{array}{ccc}
		0 & -\omega_{3i}  &  \omega_{2i} \\
		\omega_{3i} & 0  &  -\omega_{1i} \\
		-\omega_{2i} & \omega_{1i} & 0 
	\end{array}
	\right).
	\label{wmat3}
\end{equation}

Figure \ref{fig1} shows the behavior of the system for four different sets of parameters. In all cases the left panel shows the time evolution of the module, $p$, and the components, $p_1$, $p_2$ and $p_3$, of the order parameter $\vec{p}$ and the right panel shows $p_1$ versus $p_2$. We fixed $\alpha = 0.5$ in all cases.  For panels (a) to (f) we have sampled all natural frequencies $\omega_{ki}$ from a Gaussian distribution of unit width centered at zero. In (a)-(b) $a=0.1$ and $b=0.5$, so that the real eigenvector $(0,0,1)$ has the largest eigenvalue. The order parameter indeed converges to $\vec{p} = (0,0,-0.64)$. In (c)-(d) $a=1$ and $b=0.5$. In this case the complex eigenvectors $(1,\pm i,0)$ have eigenvalues with real part larger than $b$. Now the transient is dominated by rotations determined by the complex eigenvalue but the real eigenvector slowly recovers and drives the order parameter towards $\vec{p}= (0,0,-0.61)$. Notice how the module of the order parameter increases when the oscillators stop rotating and stabilize around $p_3$. In (e)-(f) $a=1$ and $b=-0.5$. In this case the real eigenvalue is negative and the rotation persists. Finally in (g)-(h) we set all the natural frequencies to zero with $a=1.0$ and $b=0.5$. Now, contrary to (c)-(d), the complex eigenvector dominates the dynamics and $p \rightarrow 1$ as predicted by the theoretical analysis. 

\begin{figure}
	\includegraphics[scale=0.29]{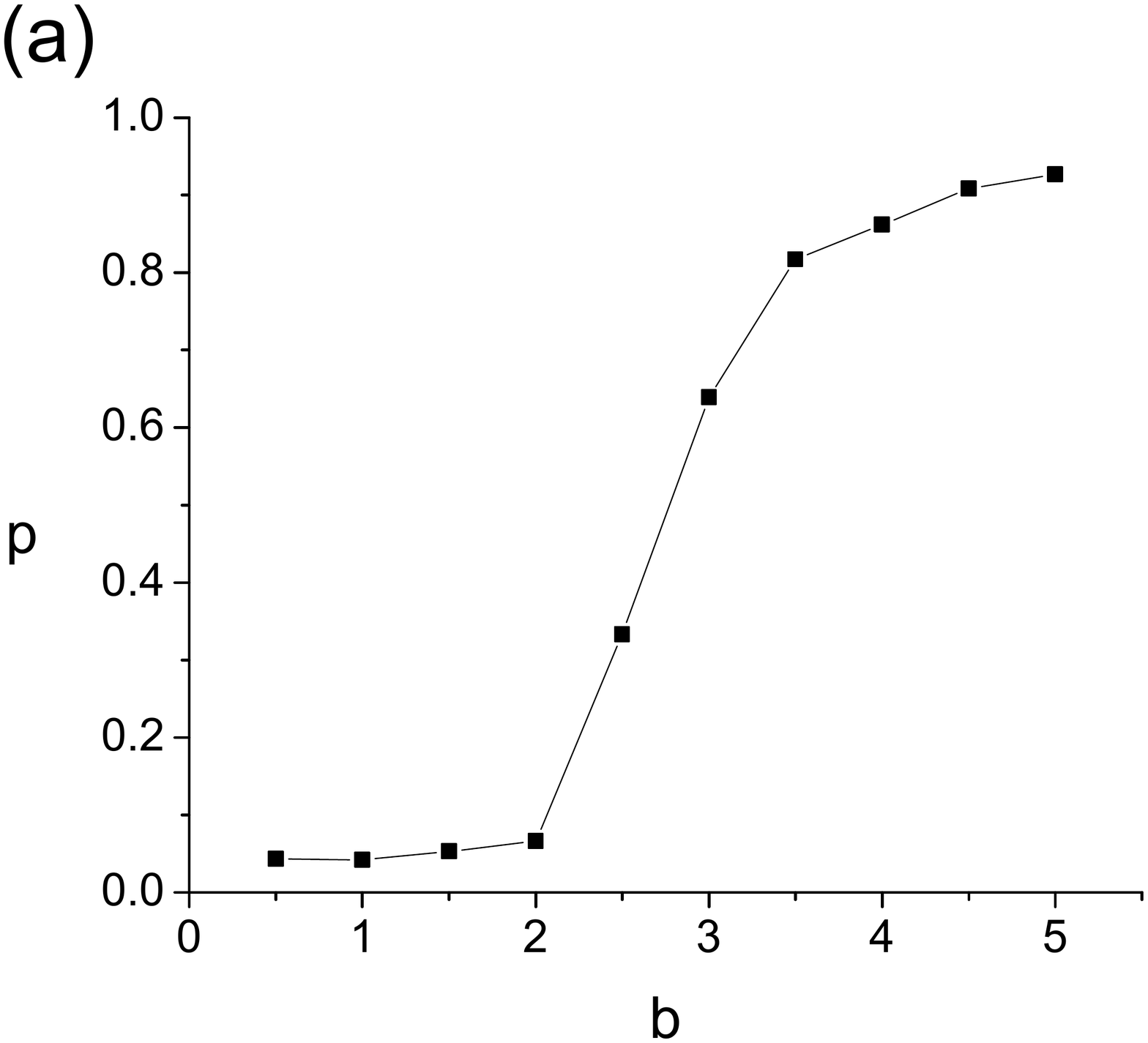} 
	\includegraphics[scale=0.29]{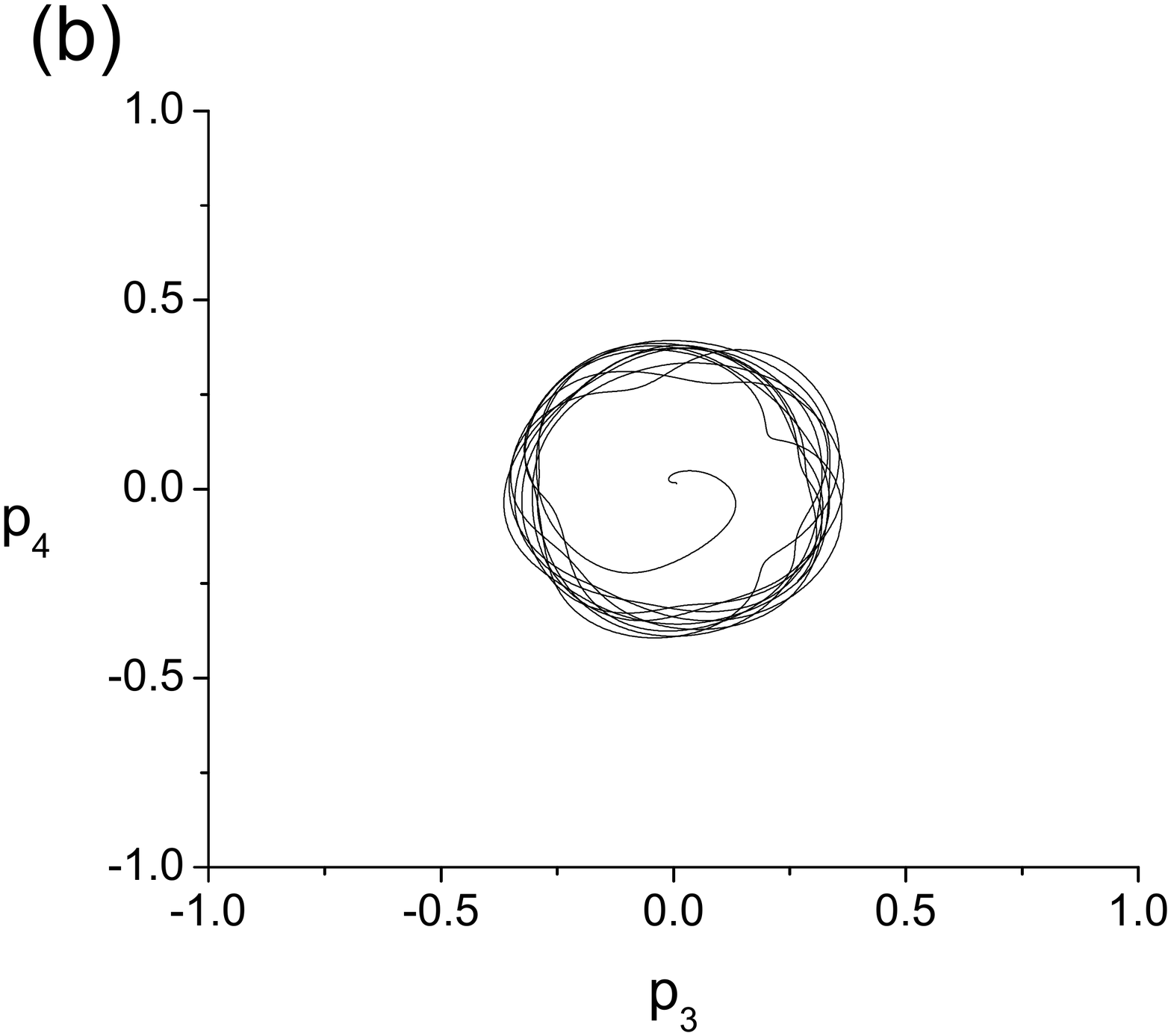} 
	\includegraphics[scale=0.29]{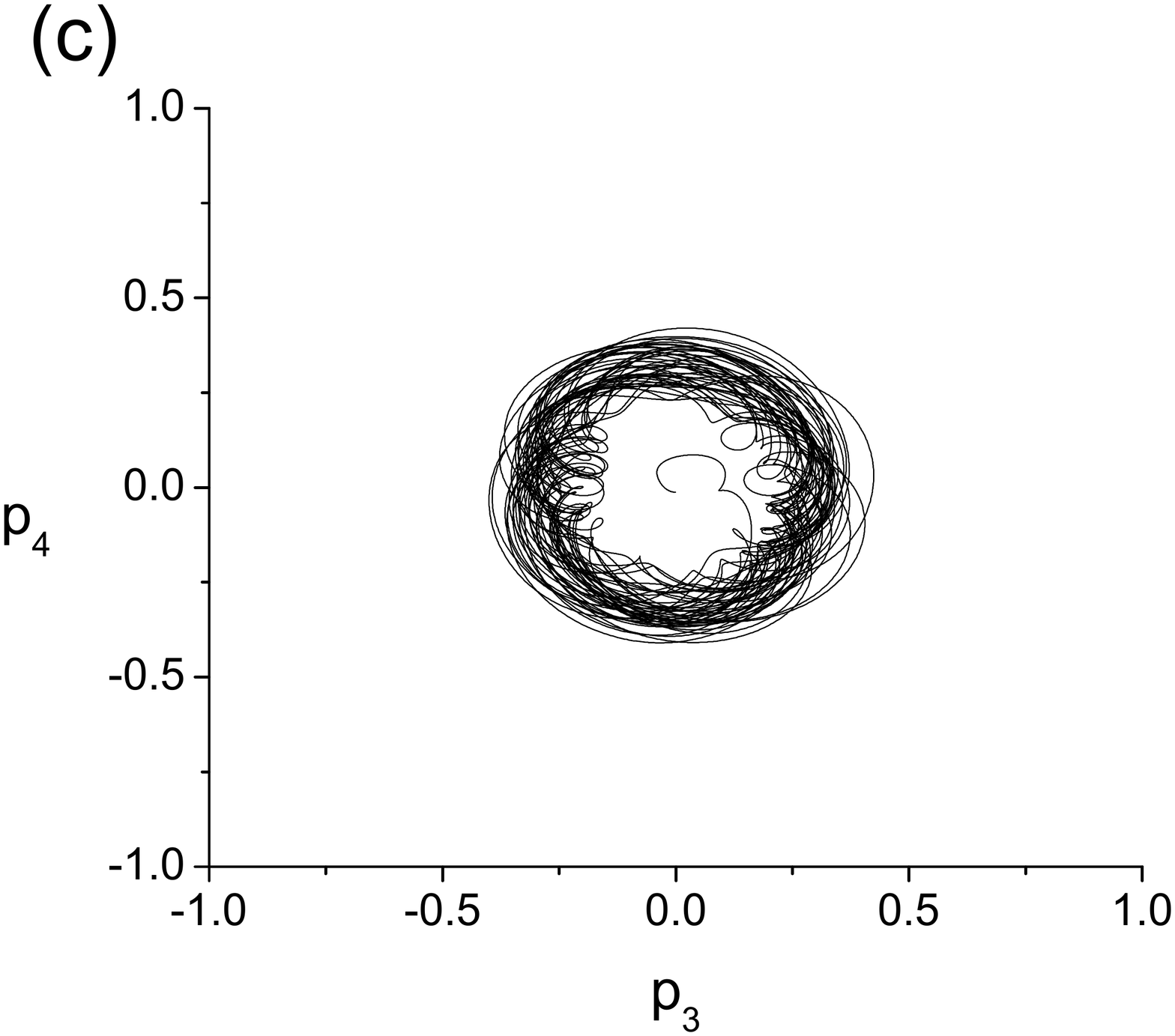} 
	\includegraphics[scale=0.29]{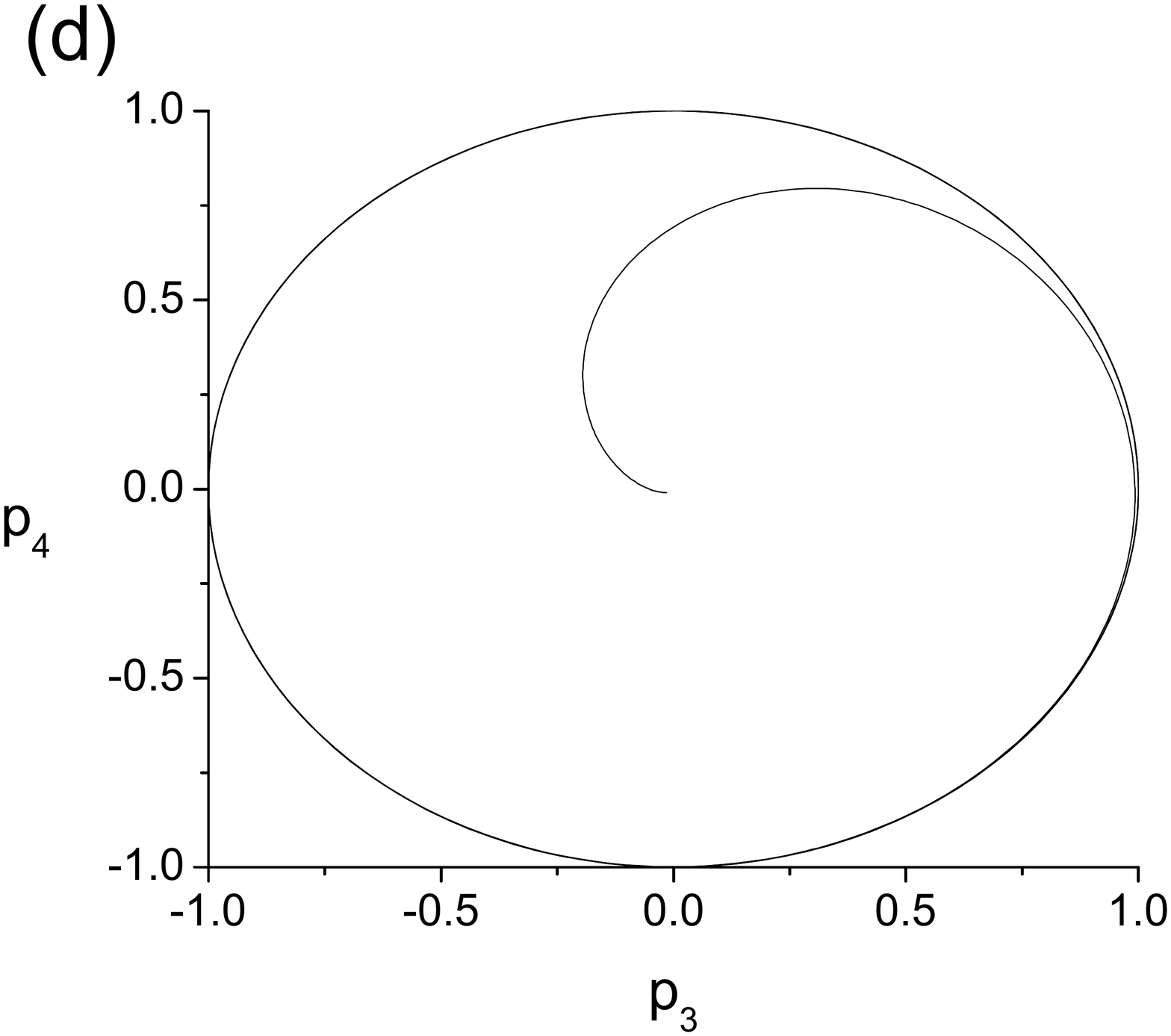} 
	\caption{4D Kuramoto model: (a) Modulus of order parameter as a function of $b$ for $\alpha=0.7$, $\beta=0.5$, $a_1=a_2=0.5$ and Gaussian distribution of natural frequencies. Panels (b) to (d) display the dynamics of order parameter in the $p_3 \times p_4$ plane for $b=2.5$ (b); for $b=2.5$ with $\alpha=0$  (c); and for $b=2.5$, $\alpha=0.5$ and zero natural frequencies (d). } 
	\label{fig2}
\end{figure}

Next we show results for $D=4$, We choose the coupling matrix as
\begin{equation}
	\mathbf{K} = \left(
	\begin{array}{cccc}
		a_1 \cos \alpha & a_1 \sin \alpha & 0 & 0\\
		-a_2 \sin \alpha & a_2 \cos \alpha & 0 & 0\\
		0 & 0 & b \cos \beta & b \sin \beta\\
		0 & 0 & -b \sin\beta & b \cos \beta 
	\end{array}
	\right),
	\label{4d}
\end{equation}
representing a rotation in the lower block and another rotation (if $a_1=a_2$) or two real eigenvectors (if $\alpha=0$ and $a_1 \neq a_2$) in the upper block. Again we set $N=1000$ oscillators.

Fig. \ref{fig2}(a) shows the modulus of order parameter as a function of $b$ for $\alpha=0.7$, $\beta=0.5$, $a_1=a_2=0.5$ and Gaussian distribution of natural frequencies. Synchronization starts close to $b=2.1$, which is close the predicted critical point in the 4D Kuramoto model with scalar coupling \cite{chandra2019continuous,barioni2021ott}. Panel (b) shows the projection of the order parameter dynamics in the $p_3 \times p_4$ plane for $b=2.5$. At integration time $t=100$ we find $\vec{p} = (-0.009,  -0.002, 0.267,  -0.243)$. Panel (c) shows the dynamics for $\alpha=0$, so that two eigenvectors of $\mathbf{K}$ are real. Contrary to the 3D case, the rotation persists. Integration time was extended to $t=500$. Finally, panel (d) is similar to (b), but with all natural frequencies set to zero. The order parameter displays a perfect rotation with module 1 as predicted by the theory.

\begin{figure}
	\includegraphics[scale=0.29]{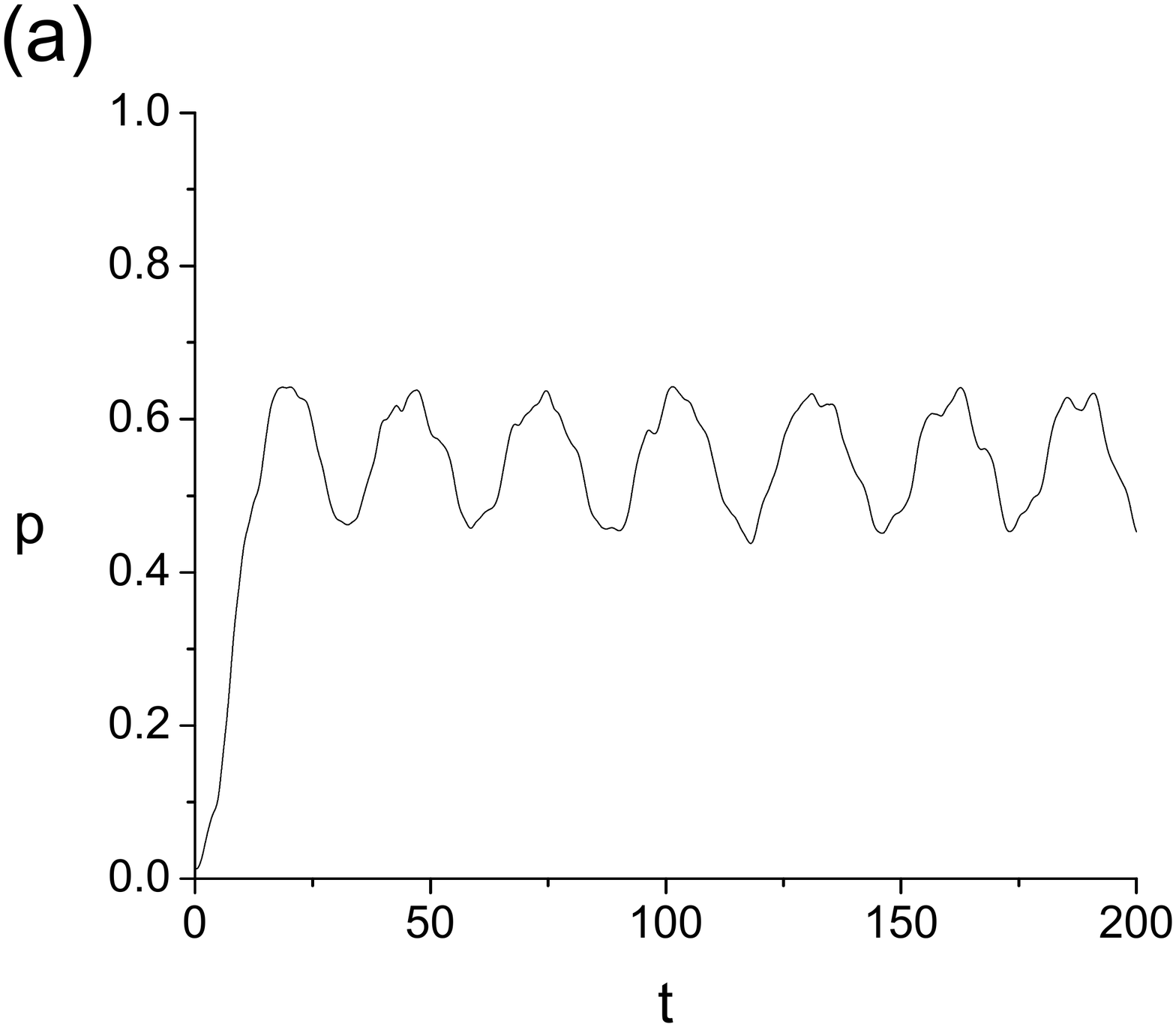} 
	\includegraphics[scale=0.29]{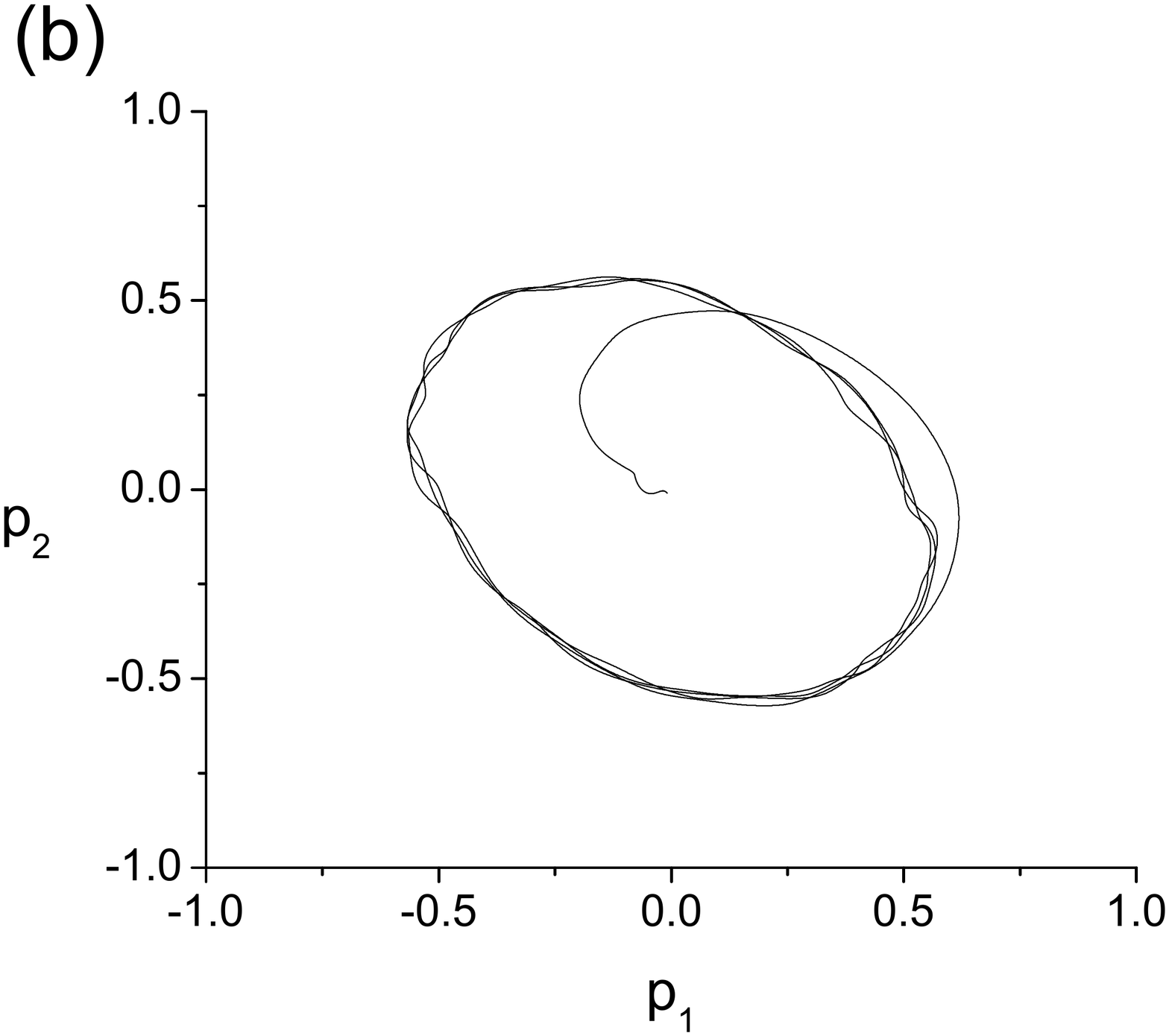} 
	\includegraphics[scale=0.29]{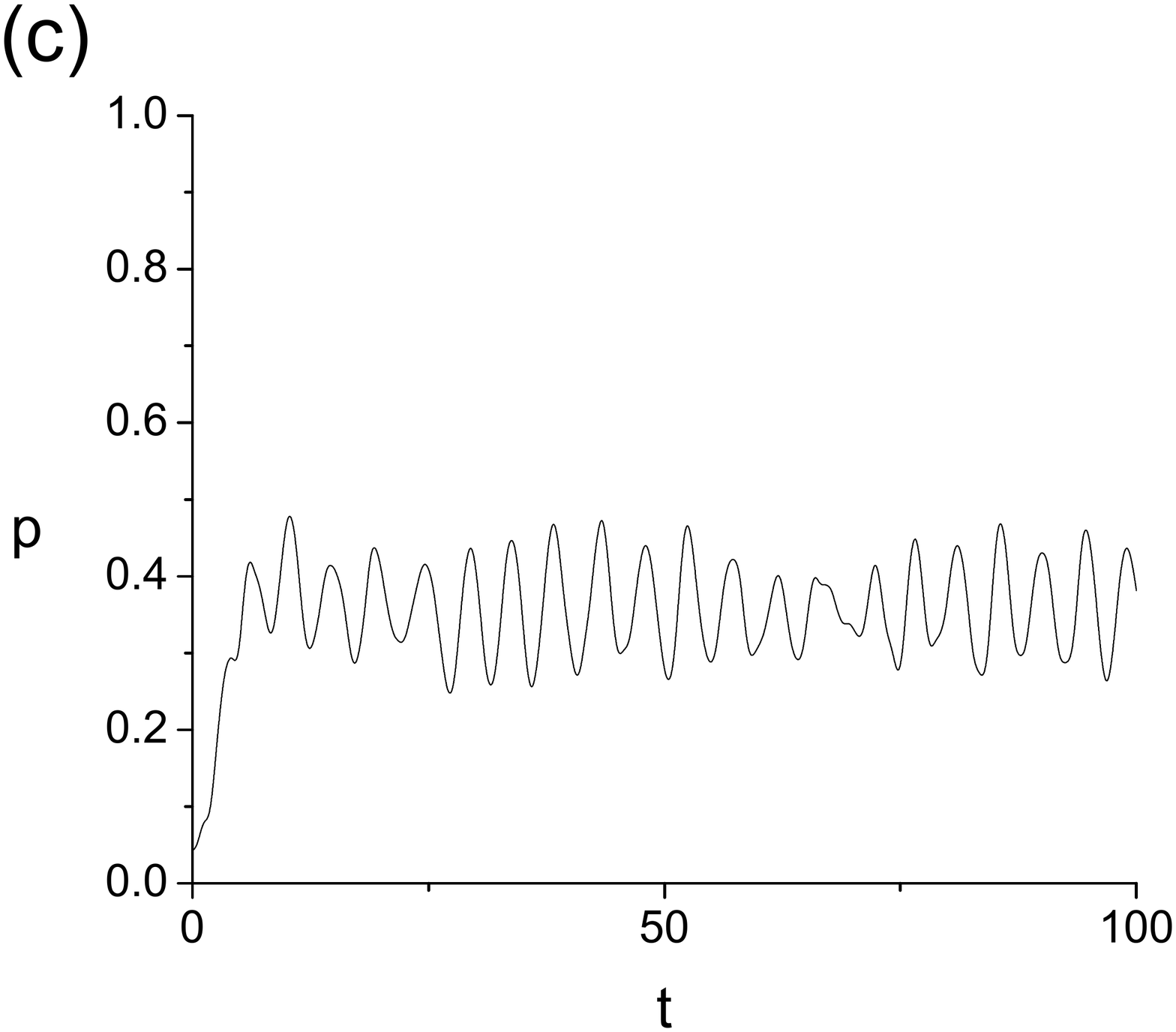} 
	\includegraphics[scale=0.29]{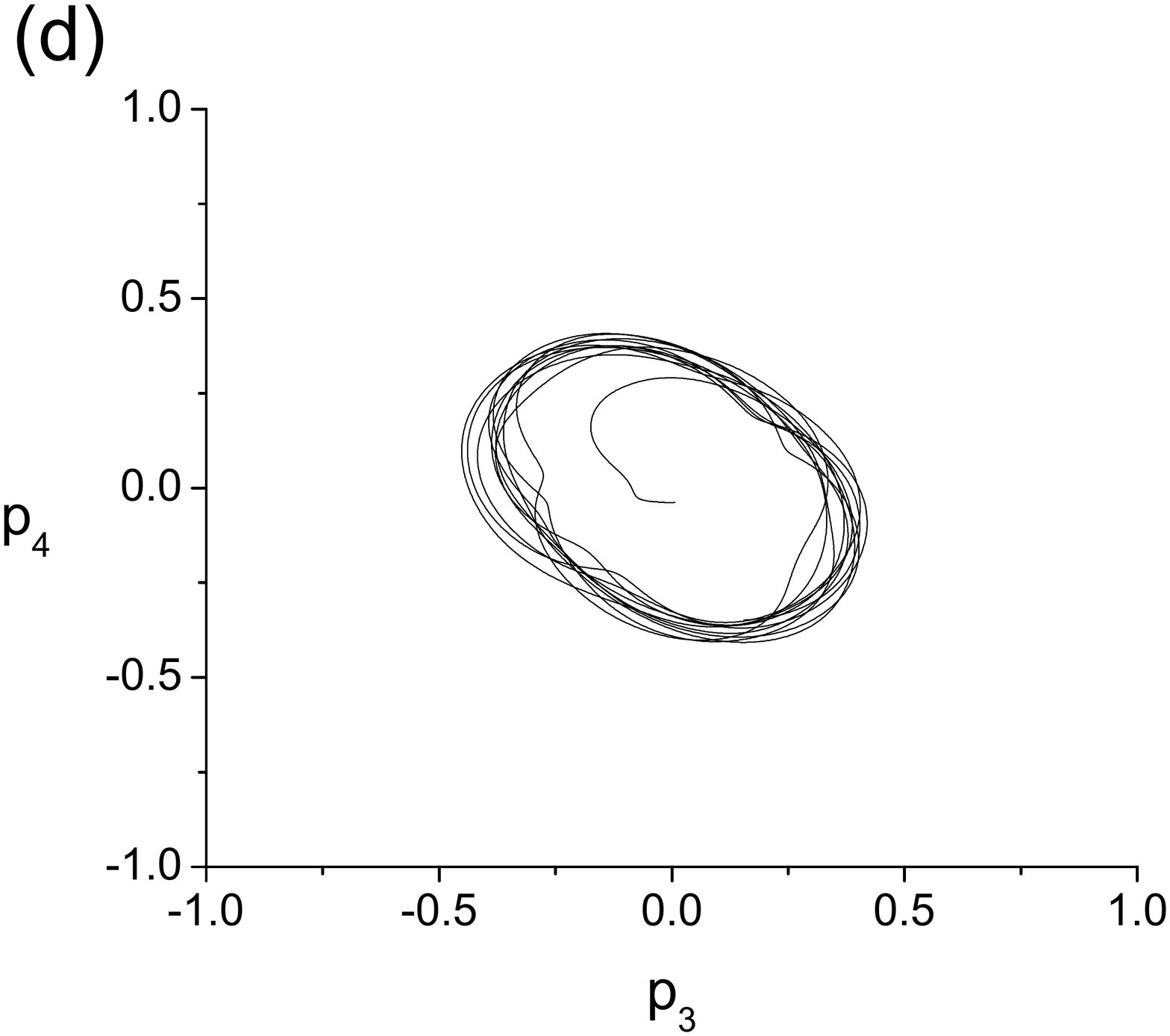} 
	\caption{Active states in 3D, ((a) and(b)) and 4D ((c) and (d)). Panels (a) and (c) show the module of the order parameter and panels (b) and (d) show its projection on the plane generated by the complex eigenvector of $\mathbf{K}$.} 
	\label{fig3}
\end{figure}

Finally we show examples of active states, where the module of the order parameter oscillates as it rotates in the plane formed by the $\vec{v}_1$ and $\vec{v}_2$, the real and imaginary parts of the complex eigenvector $\vec{u}$ of $\mathbf{K}$. As this behavior is associated with the  non-orthogonality of $\vec{v}_1$ and $\vec{v}_2$, for the 3D model we modify the coupling matrix Eq.(\ref{3d}) by setting $K_{11} = a \cos \alpha + 0.2$ and $K_{22} = a \cos \alpha - 0.2$ with $a=1$ and $b=-0.5$. The result is shown in Fig. \ref{fig3} (a) and (b). The module of $\vec{p}$ oscillates in time as the orbit in the $p_1 \times p_2$ plane traces an eliptic shape (see Eqs. (\ref{abdot1})-(\ref{eqsab})). A similar result is obtained in 4D, changing the matrix elements in Eq.(\ref{4d}) to $K_{33} = b \cos\beta + 0.4$ and $K_{44} = b \cos\beta - 0.4$ and keeping the other parameters as in Fig.\ref{fig2}(b). Figures \ref{fig3} (c) and (d) show again the module of the order parameter and its projection on the $p_3 \times p_4$ plane.

\section{Conclusions}

We have considered the $D$-dimensional version of the Kuramoto model introduced in \cite{chandra2019continuous} and \cite{barioni2021ott} where oscillators are replaced by particles moving on the surface of a sphere. The particles are described by unit vectors and are coupled by a real matrix $\mathbf{K}$ that acts on the vectors and describe a generalized form of phase frustration \cite{buzanello2022matrix}. We have shown that when the natural frequencies of all oscillators is zero (identical oscillators), the dynamics is dominated by the eigenvector of $\mathbf{K}$ with $\lambda = \lambda_M$, i.e. by the eigenvalue having the largest real part. If this eigenvalue is real and positive the oscillators converge to point on the sphere given by the direction of the corresponding eigenvector. If the eigenvalue is complex, the solution rotates in the plane defined by the real and imaginary parts of the  eigenvector.

These results change when the natural frequencies are reintroduced. For even dimensions synchronization requires a minimum value for $\lambda_M$ and there is a continuous phase transition. For odd dimensions the transition to synchronization is discontinuous and rotations are suppressed, occurring only if all other eigenvalues have negative real parts. These results were discussed only qualitatively, and a next step would be to demonstrate them analytically. It would also be important to understand how the synchronized states depend on the average value of the natural frequencies. As rotations do not generally commute with $\mathbf{K}$, changing the average of the natural frequencies is not equivalent to change reference frames (see  \cite{buzanello2022matrix} for $D=2$). Other directions in this study are to understand the response of the system to time dependent external forces \cite{moreira2019global,climaco2019optimal} and the role of coupling matrices in the dynamics of swarmalators \cite{o2017oscillators,o2022collective,Lizarraga2020}.

Our results also show  that synchronized states can be controlled by the coupling matrix. Choosing different matrices for specific  groups of particles can be a simple way  to describe different oscillatory patterns on a  modular network of connections. This, in turn, can be used to associate different functions  to the modules \cite{moreira2019modular}, whereas interactions between modules or with external sources would be responsible for the global behavior of the system. 


\begin{acknowledgments}
	It is a pleasure to thank Alberto Saa, Jose A. Brum and Guilhermo L. Buzanello for helpful suggestions. 
	This work was partly supported by FAPESP, grants 2016/01343‐7 and 2021/14335-0 (ICTP‐SAIFR) and CNPq, grant 301082/2019‐7. 
\end{acknowledgments}

\begin{appendix}
	
	\section{Rotating states}
	\label{appa}
	
	From Eqs.(\ref{eqsab}) we find
	\begin{equation}
		\dot{A} B - \dot{B} A = -\lambda_2 (A^2 + B^2)
	\end{equation}
	or
	\begin{equation}
		B^2 \frac{d}{dt}\left(\frac{A}{B} \right) = -\lambda_2 (A^2 + B^2).
	\end{equation}
	Defining $x = A/B$ this equation simplifies to $\dot{x} = -\lambda_2 (1+x^2)$ whose solution is $x = \cot (\lambda_2 t) $. Using the normalization condition 
	$A^2 (1+g_{12}) + B^2(1-g_{12}) = 1$ we arrive at the solutions (\ref{absol}).
	
	\section{Linearized equations for the active solutions}
	\label{appb}
	
	In this appendix we derive the equations describing the linearized dynamics in the neighborhood of an active state, generated by the complex eigenvalues of $\mathbf{K}$. We start with the terms on the right hand side of Eq.(\ref{eqxcpx}). Using Eqs. (\ref{v1v2}) and (\ref{kv1v2}) we compute
	\begin{equation}
		\mathbf{K} \vec{w} = [\lambda_1 - \lambda_2 g_{12}(\alpha^2-\beta^2) ] \vec{w} + \lambda_2 \eta^{-1} (\alpha^2+\beta^2) \vec{z}
	\end{equation}
	and
	\begin{equation}
		\mathbf{K} \vec{z} =  [\lambda_1 + \lambda_2  g_{12}(\alpha^2-\beta^2) ] \vec{z} - \lambda_2 \eta [1+2\alpha\beta g_{12} + g_{12}^2(\alpha^2+\beta^2)] \vec{w}.
	\end{equation}
	Then, using $\vec{w} \cdot \vec{z} = 0$,
	\begin{equation}
		(\vec{w} \cdot \mathbf{K} \vec{w}) \vec{x}_i = [\lambda_1 - \lambda_2 g_{12}(\alpha^2-\beta^2) ] \, ( a_{iz} \vec{z} + \sum_\gamma a_{i\gamma} \vec{V}_\gamma).
	\end{equation}
	Using also that $\vec{w} \cdot \vec{V}_\gamma = 0$,
	\begin{equation}
		(\vec{x}_i \cdot \mathbf{K} \vec{w}) \vec{w} = \lambda_2 g_{12} \eta^{-1} (\alpha^2+\beta^2) [a_{iz} + \sum_\gamma a_{i\gamma} g_{z\gamma}] \, \vec{w}.
	\end{equation}
	Finally we compute
	\begin{equation}
		\mathbf{K} \vec{z}-(\vec{w} \cdot \mathbf{K} \vec{z}) \vec{w} = [\lambda_1 - \lambda_2 g_{12}(\alpha^2-\beta^2) ] \vec{z}
	\end{equation}
	and
	\begin{equation}
		\mathbf{K} \vec{V}_\gamma-(\vec{w} \cdot \mathbf{K} \vec{V}_\gamma) \vec{w} = \lambda_\gamma \vec{V}_\gamma - \lambda_2 g_{w\gamma} \eta^{-1}(\alpha^2+\beta^2) ] \vec{z}
	\end{equation}
	which, together, give
	\begin{eqnarray}
		\mathbf{K} \vec{x}_j-(\vec{w} \cdot \mathbf{K} \vec{x}_j) \vec{w} 
		& = & \left\{ a_{jz}[\lambda_1 - \lambda_2 g_{12}(\alpha^2-\beta^2)] - \sum_\gamma a_{j\gamma} \lambda_2 g_{w\gamma} \eta^{-1}(\alpha^2+\beta^2)\right\} \vec{z} \\ \nonumber
		& & + \sum_\gamma a_{j\gamma} \lambda_\gamma \vec{V}_\gamma .
	\end{eqnarray}

	We now compute the left hand side of Eq.(\ref{eqxcpx}) taking into account that $\vec{w}$ and $\vec{z}$ are time-dependent vectors:
	\begin{equation}
		\dot{\vec{x}}_i = \dot{a}_{iz} \vec{z} + a_{iz} \dot{\vec{z}} + \sum_\gamma \dot{a}_{i\gamma} \vec{V}_\gamma - \sum_\gamma a_{i\gamma} \dot{g}_{w\gamma} \vec{w}- \sum_\gamma a_{i\gamma} g_{w\gamma} \dot{\vec{w}}.
	\end{equation}
	Expressions for $\dot{\vec{w}}$ and $\dot{\vec{z}}$ are given by Eqs. (\ref{wdot}) and (\ref{zdot}) respective. Also
	\begin{equation}
		\dot{g}_{w\gamma} = \dot{\vec{w}} \cdot v_\gamma = \lambda_2 \eta^{-1} (\alpha^2 + \beta^2) g_{z\gamma}.
	\end{equation}
	
	Substituting these results into Eq.(\ref{eqxcpx}) we see that all terms in $\vec{w}$ cancel out. The terms proportional to $\vec{V}_\gamma$ result in Eq.(\ref{eqgamma}). Finally, the terms proportional to $\vec{z}$ give
	\begin{eqnarray}
		\dot{a}_{iz} &-& \lambda_2 \eta^{-1} (\alpha^2 + \beta^2)\sum_\gamma a_{i\gamma} g_{w\gamma} = - a_{iz} [\lambda_1 + \lambda_2  g_{12}(\alpha^2-\beta^2) ] \\ \nonumber
		& & \frac{1}{N} \sum_j \left\{ a_{jz}[\lambda_1 - \lambda_2 g_{12}(\alpha^2-\beta^2)] - \sum_\gamma a_{j\gamma} \lambda_2 g_{w\gamma} \eta^{-1}(\alpha^2+\beta^2) \right\}.
	\end{eqnarray}
	If $0 <\lambda_1 > \lambda_\gamma$, the coefficients $a_{i\gamma}$ converge to zero and this equation simplifies to Eq. (\ref{eqzdot}):
	\begin{eqnarray}
		\dot{a}_{iz} &=& - a_{iz} [\lambda_1 - \lambda_2 g_{12}(\alpha^2-\beta^2)] + \frac{1}{N} \sum_j a_{jz} [\lambda_1 + \lambda_2 g_{12}(\alpha^2-\beta^2)].
	\end{eqnarray}
	In vector form it can also be written as
	\begin{eqnarray}
		\dot{\vec{a}}_{z} = M \vec{a}_z + N(t) \vec{a}_z
	\end{eqnarray}
	where $M = -\lambda_1 (\mathbf{1} - \frac{1}{N} \mathbf{O})$  and $N(t) = f(t) (\mathbf{1} + \frac{1}{N} \mathbf{O})$, with $f(t) = \lambda_2 g_{12}(\alpha^2-\beta^2)$ and $\mathbf{O}$ is a matrix with all entries equal to 1, $O_{ij}=1$. Because $\mathbf{M}$ and $\mathbf{N}$ commute we can define
	\begin{equation}
		\vec{b}_z = e^{-\int_0^t N(t') dt'} \vec{a}_z
	\end{equation}
	to get $   \dot{\vec{b}}_{z} = M \vec{b}_z$. Since the integral of $f(t)$ over one period is zero and the eigenvalues of $\mathbf{M}$ are $-\lambda_1$, with degeneracy (N-1) and 0, it suffices to have $\lambda_1 > 0$ to guarantee stability. The eigenvalue 0 corresponds to displace all oscillators by the same amount in the direction $\vec{z}$, which is the direction of $\dot{\vec{w}}$.
	
\end{appendix}

\clearpage 
\newpage

\end{document}